%% file: main.tex
\documentclass[12pt,preprint]{aastex631}

\input{preamble}
\begin{document}
\shorttitle{Morphological Characterization of FRBs With Deep Learning}
\shortauthors{B. Kharel et al.}
\title{Repeating versus Nonrepeating Fast Radio Bursts: A Deep Learning Approach to
Morphological Characterization}
\input{authors}

\correspondingauthor{Bikash Kharel}
\email{bk0055@mix.wvu.edu}
\submitjournal{The Astrophysical Journal}

\begin{abstract}
 We present a deep learning approach to classify fast radio bursts (FRBs) based purely on morphology as encoded on recorded dynamic spectrum from \cfrb \cattwo. We implemented transfer learning with a pretrained \convnext architecture, exploiting its powerful feature extraction ability. \convnext was adapted to classify dedispersed dynamic spectra (which we treat as images) of the FRBs into one of the two sub-classes, i.e., repeater and non-repeater, based on their various temporal and spectral properties and relation between the sub-pulse structures. Additionally, we also used mathematical model representation of the total intensity data to interpret the deep learning model. Upon fine-tuning the pretrained \convnext on the FRB spectrograms, we were able to achieve high classification metrics while substantially reducing training time and computing power as compared to training a deep learning model from scratch with random weights and biases without any feature extraction ability. \revieweredit{Importantly, our results suggest that the morphological differences between CHIME repeating and non-repeating events persist in \cattwo and the deep learning model leveraged these differences for classification.} The fine-tuned deep learning model can be used for inference, which enables us to predict whether an FRB'                    s morphology resembles that of repeaters or non-repeaters. Such inferences may become increasingly significant when trained on larger data sets that will exist in the near future. 

\end{abstract}


\section{Introduction}
\label{sec:intro}
In recent years, machine learning has seen widespread application in almost every field of astrophysics. Clustering \citep[e.g.,][]{Banerjee2023}, dimensionality reduction \citep[e.g.,][]{ZhuGe2022}, neural network models such as convolutional neural networks (CNNs) developed for imaging \citep[e.g.,][]{Agarwal2020}, regression, and classifier \citep[e.g.,][]{Ness2015} models are the most common machine learning techniques implemented to date. In particular, classification tasks have been used to perform automated morphological classification for large collection of galaxy images \citep[e.g.,][]{Dieleman2015}, transport models for the active galactic nuclei (AGN) \citep[e.g.,][]{Sanchez-Saez2021}, variability classification of transients, redshift estimation for AGN and GRB events \citep[e.g.,][]{Pasquet-Itam2018}, and detection of repeating FRBs \citep[e.g.,][]{Zhang_2018}.

Fast radio bursts (FRBs) \referee{\citep{doi:10.1126/science.1147532}} are bright millisecond radio pulses of mostly extra-galactic origin \citep{Petroff2022}; with only one in the Milky Way Galaxy has been found so far \citep{Bochenek2020, CHIMEFRB_Collaboration2020-vf}. The underlying physics governing the emission mechanism of such intense radio signals is still uncertain. Several FRB emission models have been proposed, including but not limited to \revieweredit{magnetospheric curvature emission \citep{Wang__2022}}, synchrotron maser emissions \citep{10.1093/mnras/stz700, 10.1093/mnras/stz640}, \revieweredit{magnetospheric emission \citep{Nimmo2025-vi}}, and stimulated emission \citep{Dogan2020}. However, none of the aforementioned models totally explain all FRB properties. Intriguingly, some FRBs are observed to be repeating over time \citep{spitler2016repeating,Scholz2016}, while others apparently remain dormant after emitting a single burst, \revieweredit{suggesting} two distinct classes of bursts. Different machine learning techniques \citep[e.g.,][]{JUNIOR2026100449, Luo_2022, Sun_2025,sun2025practicalframeworkestimatingrepetition, madheshwaran2025machinelearningclassificationbaseband} have been applied so far in an attempt to classify the FRBs into these two subclasses. However, most of these methods have been limited by one or more factors such as small FRB sample, imbalanced data set during training and evaluation, feature expansion or aggregation, or mixing of non-morphological \revieweredit{and} morphological features. \referee{Only few analysis relied on purely morphology based studies using dynamic spectrum \citep[e.g.,][]{Yang_2023, 2025MNRAS.538..408K}}.

The Canadian Hydrogen Intensity Mapping Experiment (CHIME) telescope \citep{CHIME2022} is a \revieweredit{transit} radio interferometer operating in the 400–800 MHz range. CHIME uses a commensal-science approach to make daily, simultaneous measurements for 21-cm cosmology, radio pulsars, and FRBs. The \cfrb project detects 2-3 FRBs per day, observing the sky north of declination -11 degrees every day with \revieweredit{an instantaneous} field of view of $\sim$200 degrees. \cfrb \catone \citep{CHIME2021} consisted of 536 events, of which 62 were from 18 previously reported \revieweredit{repetitive} sources. The \catone data set was evaluated statistically by  \citep{Pleunis2021}, \revieweredit{who discerned} morphologically contrasting behavior of repeating and apparently non-repeating FRBs, thereby suggesting different  emission mechanisms and local environments \revieweredit{for} the two sub-populations. 

The number of FRBs reported by CHIME since the release of \catone has increased approximately eightfold, along with a significant rise in the number of repeating FRBs both from the sources in \catone  and from new sources. \cfrb \cattwo \citep{chime_cat_2} includes \nfrbtot FRBs, \revieweredit{comprising} \nfrbnorep  one-off events and \nfrbrep repeating bursts from \nrepeater different sources. This surge in the number of FRBs detected by CHIME in \cattwo provides an unprecedented opportunity to further investigate the morphological characteristics associated with \revieweredit{the} two \revieweredit{potential}  sub-classes of the FRBs. Using machine learning, we are not only able to observe the morphological differences but also can make predictions for new FRBs,  which could be very useful for follow up observations and other analyses. 

This article explores into the purely morphological classification between repeating and seemingly non-repeating FRBs from CHIME \cattwo with machine learning, using image pattern recognition. \revieweredit{This research is important as it sheds light on the pure morphological characteristics of \cfrb total intensity data set and the potential for using machine learning tools to analyze them}.  We describe the data sets used in this analysis in Section \ref{sec:chime and frb data}. In Section \ref{sec:CNN Model For chime-frb images}, we outline the deep neural network (DNN) model architecture, transfer learning and data preprocessing steps implemented for our analysis. Section \ref{sec: training_real} explains the fine-tuned \convnext model \revieweredit{applied to}  \cfrb \cattwo total intensity data and provides results and interpretations. Section \ref{sec:fine tuning on fitburst modeled data} consists of fine tuning the \convnext model on \fitburst generated images and interpreting this fine-tuned model using synthetic \fitburst generated images. Finally, the Section \ref{sec:discussions and conclusions} provide  discussions and conclusions from this work.  

\section{Images from \cfrb Data}
\label{sec:chime and frb data}
The CHIME telescope \citep{CHIME2022} consists of four 20m x 100m cylindrical reflectors that are equipped with 1,024 dual-polarization feeds. The feeds are tuned to receive radiation in the 400--800 MHz range and are connected to an ``FX" correlator powered by graphics processing units (GPUs), which form 1,024 beams of timeseries on the sky. For FRB detection, the raw voltage data from each feed are added in qaudrature to form total intensity timeseries, up-channelized into $16, 384$ frequency bins, and downsampled to 0.983-ms time resolution.\footnote{While CHIME initially generates data across 1,024 frequency channels, the upchannelization by 16 times is implemented to minimize intrachannel smearing of pulses with \revieweredit{large} dispersion measures \revieweredit{(DMs), e.g., DM} $>$ 1000 pc cm$^{-3}$.} These dynamic spectra are then searched for FRBs by the \cfrb backend using a multistage real-time process \citep{Amiri_2018}.  

Upon the identification of an FRB candidate, the \cfrb system stores several seconds of full-resolution total intensity data along with metadata including timestamps, beam coordinates, and radio frequency interference (RFI) flags. The \fitburst framework \citep{Fonseca2024} is then employed to measure burst properties like width, spectral index, flux, and scattering through direct modeling of pulse morphology. When executed on recorded \cfrb data, \fitburst extracts a dedispersed segment of the dynamic spectrum using the best available information of \revieweredit{dispersion measure (DM)} and time of arrival; these ``windowed" data typically span 162 time samples, with the burst at the middle of the window, but can be larger if the burst exhibits a high degree of structure (e.g., large scatter-broadening, multiple components, etc.). 

\begin{figure}[h!]
\centering
\includegraphics[width=1\linewidth]{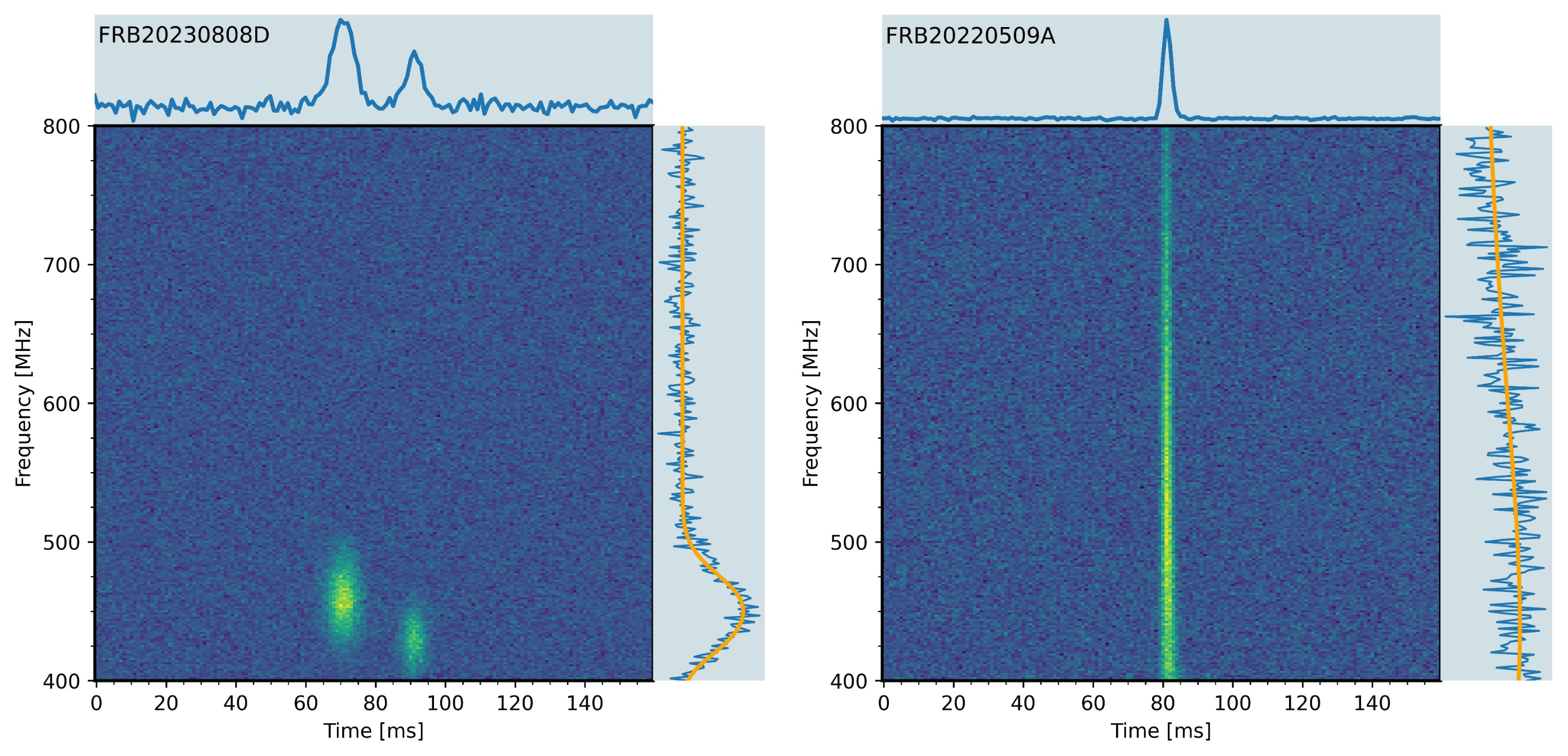}
\caption{\referee{Typical dynamic spectra of FRBs showing different burst morphology in different subclasses. (Left) A repeating FRB with narrow bandwidth and broader pulse width. (Right) An apparently non-repeating broad-band FRB with narrow pulse width. Any other bursts from the source haven't been recorded to date.}}
\label{fig:real_dynamic_spectra}
\end{figure}

For this work, we use results from the second \cfrb catalog and the windowed data subject to modeling as obtained by \fitburst. These windowed dynamic spectra are hereafter referred to as ``images" and form the foundation of our analysis, along with the best-fit \fitburst parameters published in \cattwo.
Figure \ref{fig:real_dynamic_spectra} displays two typical FRB images of seemingly different burst morphologies, from a repeating and a non-repeating events. 

\section{A CNN Model for \cfrb Images}
\label{sec:CNN Model For chime-frb images}
\begin{figure}[ht!]
  \centering 
\includegraphics[width=1\textwidth]{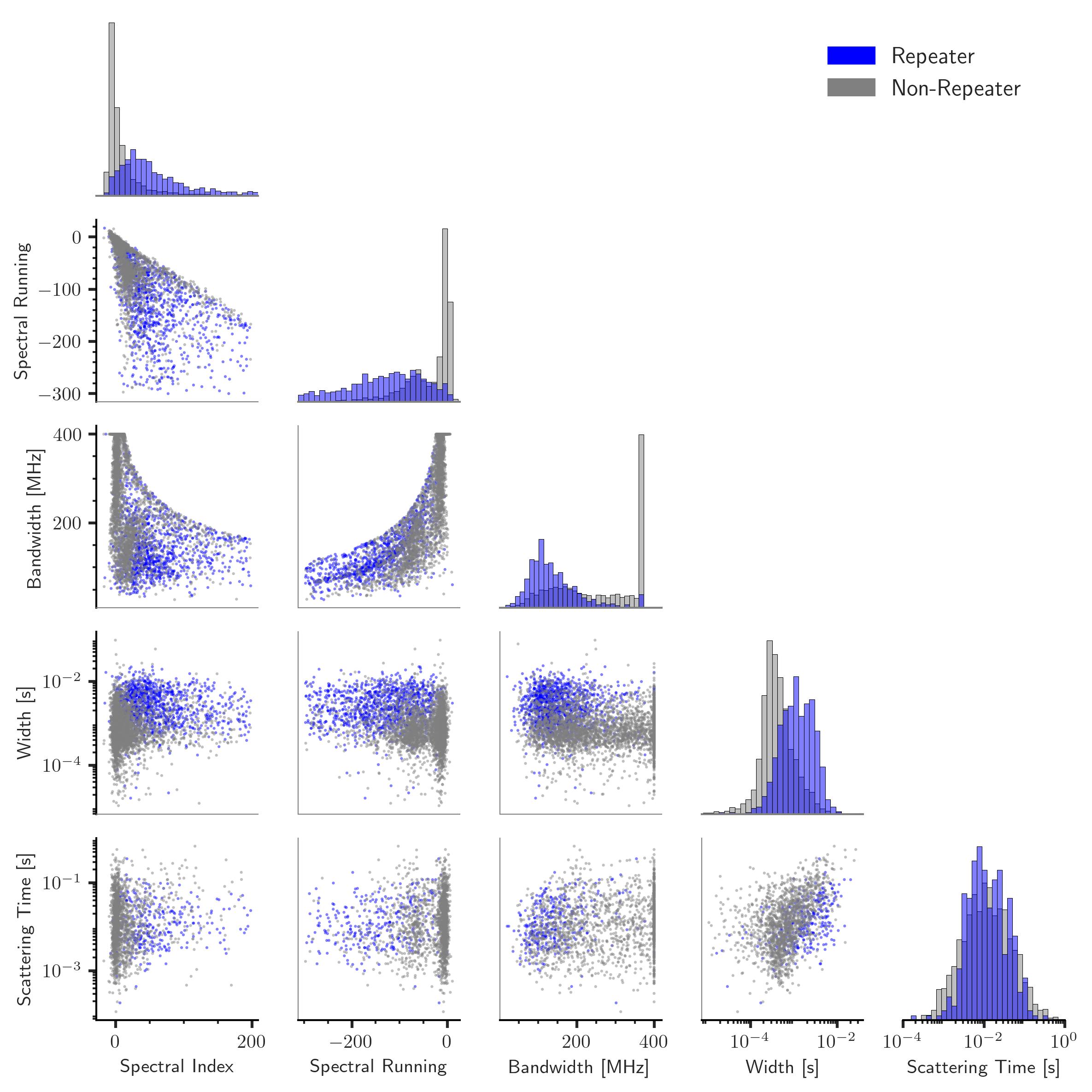} 
  \caption{\revieweredit{Corner plot for different morphological parameters in \cfrb \cattwo. Blue denotes repeating FRBs and gray denotes as yet non-repeating FRBs.}}
  \label{fig:traingle_plot}
\end{figure}

Figure \ref{fig:traingle_plot} displays a corner plot for the \fitburst-measured morphological parameters of \cfrb \cattwo events with feature expansion (i.e. treating each component as an individual burst) for multi-component bursts. \revieweredit{Each blue dot represents either a repeating event or a component from a repeating multi-component event. Similarly, each gray dot represents a non-repeating event or a component from multi-component non-repeating burst.}
The plot clearly demonstrates the morphological differences between the repeating and apparently non-repeating FRBs, as well as correlations between the parameters. \referee{A non-trivial overlap between between the distributions of the parameters of the two sub-classes can also be observed.} The large and growing number of measurements can naturally be evaluated using certain statistical and/or machine learning tools.
\label{sec:characterization}
The various methods to examine the morphological distinctions among FRB sub-populations include but are not limited to: parameter-based classical statistical tests such as those used to study the morphological properties of \catone \citep{Pleunis2021}, parameter based classical machine learning approaches \citep[e.g.][]{luo2023machine}, parameter-based multilayer perceptron (MLP), and image-based convolutional neural networks (CNNs)\footnote{\revieweredit{Besides CNN, we also experimented with various other state of the art computer vision Deep Learning architectures like Vision Transformer, Shifted Window Transformer, but those resulted in poor evaluation metrics.}} \revieweredit{explained in Appendix \ref{appendix:cnn}}.  These methods are important for FRB science, as a significant portion of the FRB population consists of bursts with complex morphologies, such as multiple components in a burst and the frequency drift of sub-bursts over time \citep[e.g.,][]{Hessels2019}. 

Traditional statistical techniques, classic machine learning approaches and MLP models require input in the form of a single data point in $\textbf{R}^n$ spanned by $n$ morphological parameters. However, for bursts with multiple peaks, it becomes necessary to perform either feature aggregation or feature expansion for the aforementioned methods, which may introduce biases \citep{bonetti2023nonlinearfeatureaggregationalgorithms} and fail to capture the correlation between complex morphologies and sub-classes of FRBs. Additionally, using statistical learning has a major drawback in that it is unable to make inferences about newly detected FRBs \revieweredit{when there is a significant overlap between the distributions of measured parameters for two classes.}

By the nature of their design, CNNs can offer a significant advantage as they can be employed to discern patterns and correlations that may not be \revieweredit{able to be} detected by traditional statistical methods and MLPs. Given the superiority of CNNs over traditional statistical methods and MLPs in extracting features from complex FRB morphologies, we employ a CNN model to analyze the morphological properties of the \cattwo. 

\revieweredit{We  utilized CNN-based} \convnext architecture for feature extraction on our FRB image data set due to its simple\footnote{\convnext is often referred to as simple as compared to transformer-based vision models, as its  design and structure contain only traditional CNN blocks.} and efficient performance. \revieweredit{The details of the \convnext architecture that we used in this analysis is outlined in Appendix \ref{appendix:convnext architecture}. Instead of training the deep learning model from scratch, which would otherwise require large training data set and massive computing power, we implemented transfer learning and fine tuned the \convnext model weights on our \cfrb \cattwo data set. The details of the transfer learning and pre-trained \convnext model is provided in Appendix \ref{appendix:transfer learning}.}

\subsection{Fine Tuning on Total Intensity Data} \label{subsec:fine_tune_real}
All prior studies that used statistical or machine-learning methods for FRB classification have been performed on parameter data sets obtained by mathematical  modeling of FRBs by using some modeling framework \citep{Sun_2025, curtin2024morphology32repeatingfast, HerreraMartin2025a, Zhu_Ge_2022}. These analyses relied on a deterministic model of the pulse shape to obtain parameters, which is both time consuming and potentially suboptimal if all significant features are not adequately modeled. \revieweredit{Also, the studies employed either feature aggregation (e.g. averaging over sub-bursts' parameter values) or feature expansion (e.g. treating each sub-burst parameter as an individual FRB) for FRBs with sub-bursts thereby losing complete information encoded in a single event.} Moreover, a general mathematical model might not fully capture all the features of an individual FRB given the amount of variability in its morphological features.

Due to the aforementioned limitations of using model-dependent parameters, we trained a machine learning model for morphological characterization of FRBs that uses images of total intensity data from \cfrb \cattwo events. We expect our CNN-based model to be an improvement over prior work, as it directly analyzes total-intensity images and not \fitburst parameters. The total intensity images are obtained by the methods mentioned in Section \ref{sec:chime and frb data}. 

The following subsections cover data preparation, training outcomes, model interpretability, and result discussions, as well as the various downstream applications for the trained model.

\subsection{Data Preparation and Preprocessing} 
\label{subsec:real_data_preparation}
We preprocessed the \cattwo data described in Section \ref{sec:chime and frb data} by down-sampling frequency channels by a factor of 64 while retaining time resolution. The images were then readjusted by using bilinear interpolation to the standard size of 224 x 224 required for the ConvNext architecture. Since the dynamic spectrum is in gray-scale, we converted it to RGB in order to meet the input channel requirements of the standard ConvNext model. This conversion was achieved by setting the red, green, and blue channels to have the same pixel values as the original grayscale pixels. The pixel values were adjusted to have mean values of 0.485, 0.456, and 0.406 and standard deviations of 0.229, 0.224, and 0.225 along the red, green, and blue channels, respectively. This normalization is consistent with the pretraining performed on the ImageNet data set \citep{Woo2023ConvNeXtV2}.

\begin{figure}
    \centering
    \includegraphics[width=0.7\linewidth]{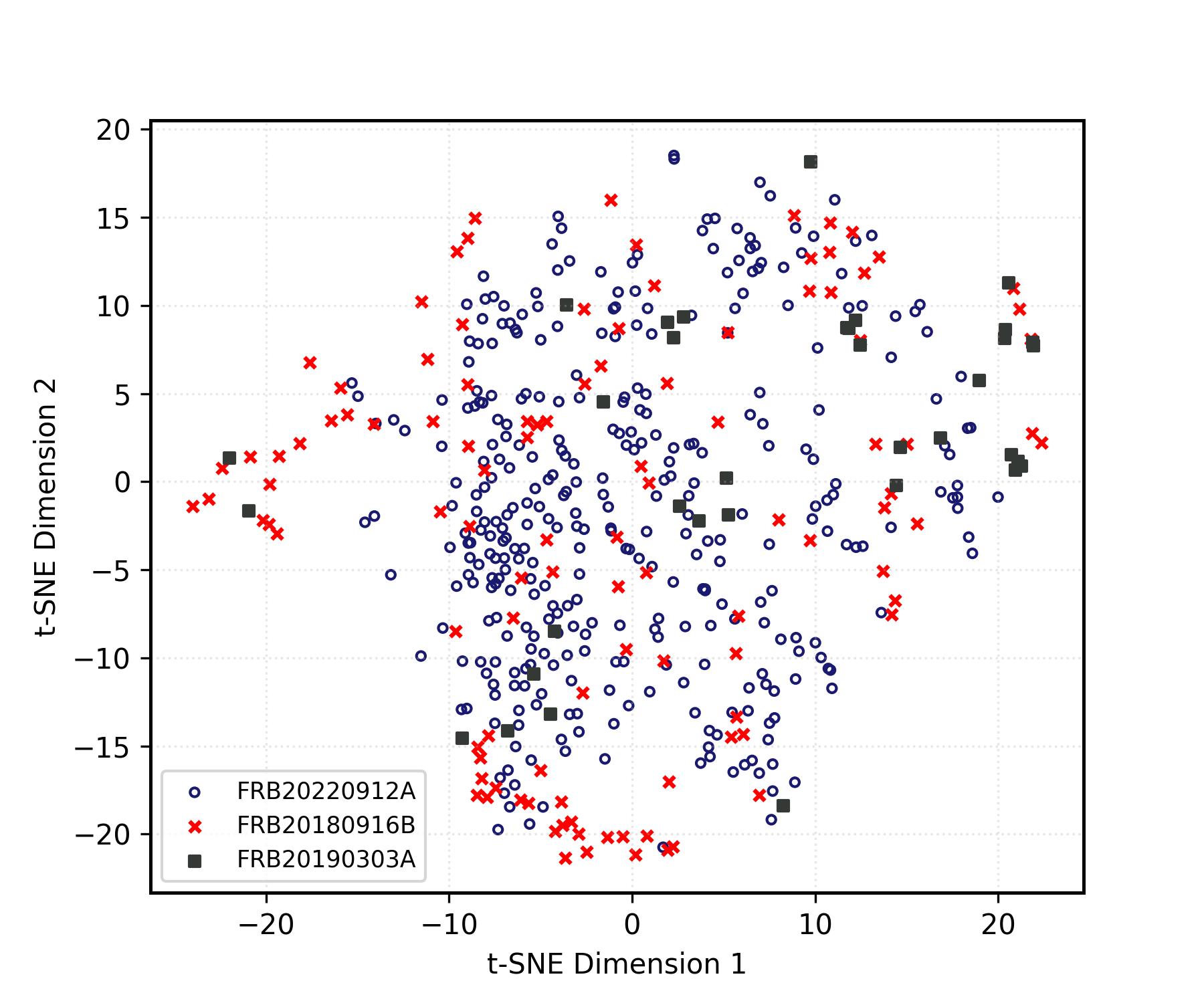}
    \caption{Distribution of latent-space features of burst morphologies for three CHIME/FRB repeating sources with the largest number of events, extracted using an autoencoder. The latent features were reduced to two dimensions with t-SNE.}
    \label{fig:repeting frbs clustering}
\end{figure}

The entire \cattwo data set was divided into three sections; train, validation, and test. \revieweredit{Our choice of selecting random events from all repeating sources for training, validation and testing was motivated by the fact that the repeating FRBs exhibit broadly similar morphological characteristics, meaning that the morphology distributions are not unique to individual repeating sources. This was verified by employing a simple unsupervised autoencoder \citep{bank2023autoencoders} approach to extract burst features. The resulting latent representations were reduced to two dimensions using t-distributed stochastic neighbor embedding (t-SNE) \citep{maaten2008visualizing}. The reduced latent space showed no evidence of clustering as shown in Figure \ref{fig:repeting frbs clustering}, suggesting no significant morphological distinctions among repeating sources. \referee{The absence of clear clustering in Figure 3 is not an artifact of the choice of hyperparameter (perplexity) in t-SNE. We varied the perplexity parameter from $30 - 100$ and observed the similar results. However, the clustering observed in previous analyses \citep[e.g.,][]{10.1093/mnras/stad1942, 10.1093/mnras/stad930} are due to the inclusion of both morphological and non-morphological features in the clustering algorithms.}} 

To fine tune pretrained \convnext model, we randomly selected 3349 of the one-off events and 673 repeating events from all of the repeating sources. Additionally, 125 samples from each class were randomly chosen for validation and testing purposes. To mitigate the issue of class imbalance during training, no synthetic data generation was carried out on the minority class, as it may lead to potential artifact biases like overgeneralization, boundary overlap and distributional mismatch \citep{Chawla_2002, 5128907}  in the training data set. We instead employed a weighted random sampling algorithm with replacement, with the weight of each class being reciprocal of its size to ensure that each class was represented adequately during the fine tuning process. The class sizes for the validation and test data sets were also made equal to avoid bias due to under-representation during model evaluation. 

\section{Analysis and Results} \label{sec: training_real}
We fine-tuned the \convnext{} model described in Appendix \ref{appendix:convnext architecture} using the high-performance computing (HPC) cluster Dolly Sods\footnote{\href{https://docs.hpc.wvu.edu/text/84.DollySods.html}{Dolly Sods documentation}.} at West Virginia University, with Graphics Processing Unit (GPU) support. All weights in the intermediate stages of the pretrained ConvNeXt were frozen, and only the weights in the ``patchify'' and fully connected layers were fine-tuned. The output layer was modified to give a single output through the sigmoid layer\citep{Dubey2021ActivationFI}, and the threshold value for classification during fine-tuning was set to 0.5. Early stopping\footnote{A regularization technique that halts training if the validation metrics (e.g., accuracy and loss) fail to improve for a specified number of iterations (the patience).} was implemented to avoid overfitting, and the learning rate was scheduled\footnote{The learning rate was dynamically adjusted during fine-tuning.} to improve model performance. The hyperparameters used in fine-tuning the pretrained \convnext{} model are listed in Table \ref{tab:hyperparameters}.

\begin{deluxetable}{l c l}
\tablecaption{Hyperparameters Used in Fine Tuning the pretrained \convnext{} Model}
\tabletypesize{\footnotesize}
\tablewidth{0pt}
\tablehead{
  \colhead{Hyperparameter} & \colhead{Value} & \colhead{Description}
}
\startdata
Learning Rate            & $1.0\times10^{-3}$ & \descbox{Initial learning rate} \\
Batch Size               & 32                 & \descbox{Number of samples per batch} \\
Maximum Epochs           & 300                & \descbox{Number of epochs set if no early stopping is implemented} \\
Learning Rate Scheduler  & 15, 30, 50, 65, 80, 100 & \descbox{Epochs at which the learning rate changes} \\
Patience                 & 20                 & \descbox{Maximum number of epochs before implementing early stopping} \\
$\Gamma$                 & 0.1                & \descbox{Reduction factor in learning rate} \\
Optimizer                & Adam               & \descbox{Algorithm to adjust parameters} \\
Activation Function      & Sigmoid            & \descbox{Activation function at the output} \\
Loss                     & Focal Loss         & \descbox{Objective function to be optimized} \\
\enddata
\tablecomments{This table summarizes the hyperparameters used during fine tuning. Other hyperparameters such as dropout rate, regularization, and hidden-layer activations were not modified from the pretrained model.}
\label{tab:hyperparameters}
\end{deluxetable}

We found that the ConvNext model underperformed in classification success when using the standard binary cross entropy as the loss function. After experimentation, we chose to adopt the focal loss \citep{lin2018focallossdenseobject} as the loss function of our ConvNext model and experimented with different focusing parameter ($\gamma$) values,  
\begin{equation}
    \text{Focal Loss} = -(1-p_t)^{\gamma}\log({p_t}),
\end{equation}
where
\begin{itemize}
    \item $\gamma$ = focusing parameter, and
    \item $p_t$ = predicted probability for the true class
\end{itemize}

\begin{figure}
    \centering
    \includegraphics[width=0.45\linewidth, height=7cm]{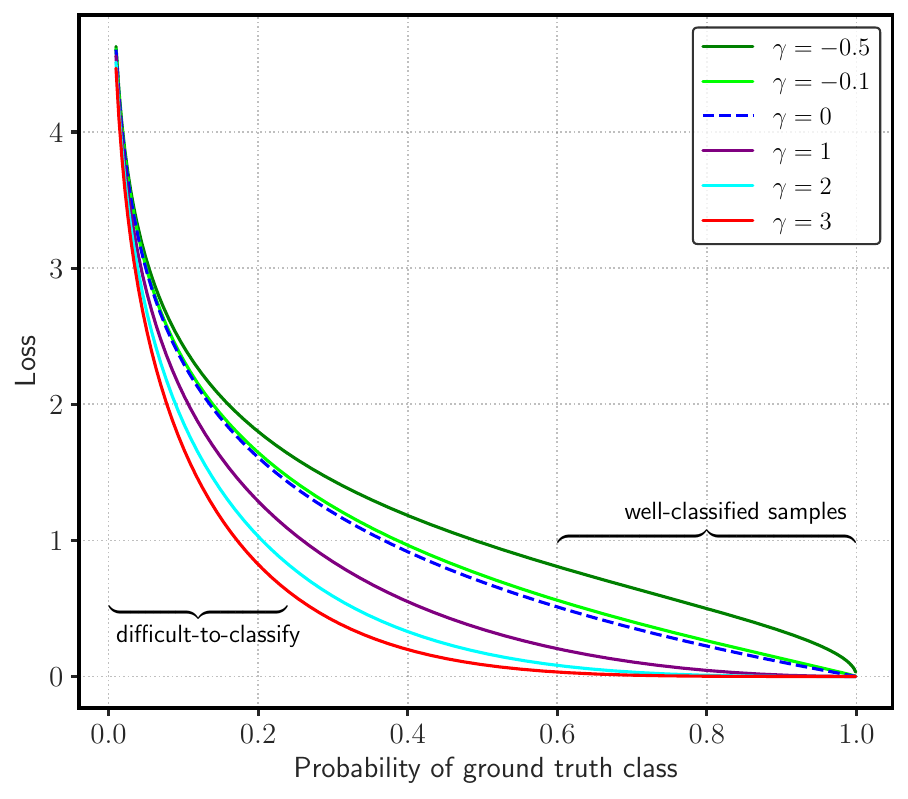}
    \hfil
    \includegraphics[width=0.45\linewidth, height=7cm]{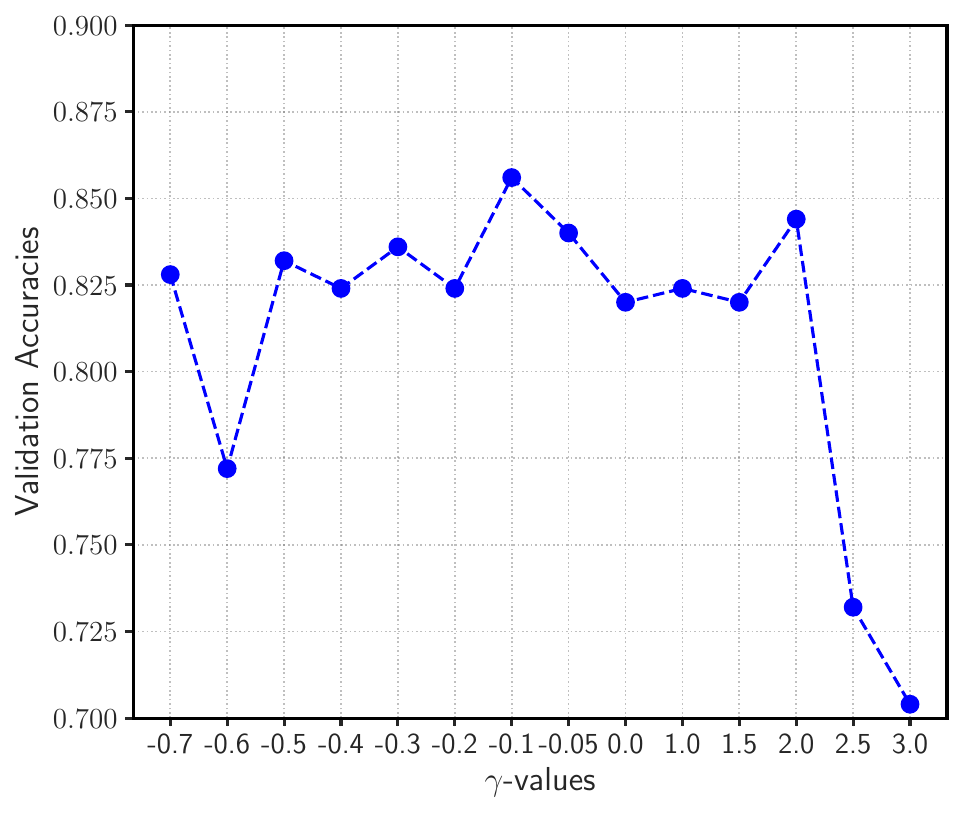}
    \caption{(Left) Plot displaying loss variations for different values of the focusing parameter in focal loss. (Right) Best validation accuracies of our fine tuned model on the validation data set.}
    \label{fig:gamma vs loss and acc}
\end{figure}

The value of $\gamma$ is usually chosen to be $\gamma \geq 0$, with $\gamma=0$ being the standard binary cross entropy loss \citep{lin2018focallossdenseobject}. Although most of the classification tasks in the past used the values of $\gamma>0$, prioritizing harder to classify samples, in  this work we also experimented with $\gamma < 0$ values, which gives more preference to the easier to classify data. Figure \ref{fig:gamma vs loss and acc} (Left) displays how a model focuses more on hard to classify samples for larger $\gamma$ values and on easy to classify samples for lower $\gamma$ values. The dashed blue line is the standard binary cross entropy loss. 
The plot of validation accuracy versus different $\gamma$ values is shown in Figure \ref{fig:gamma vs loss and acc} (Right), displaying the best accuracy when $\gamma$ is $-0.1$. The second best accuracy is obtained when the value of $\gamma$ is 2 as observed by \cite{lin2018focallossdenseobject}. Clearly the $\gamma=0$ model is under-performing as compared to models with some other $\gamma$ values. We evaluated other performance metrics as well for the two best performing models on validation data set. These metrics were calculated on the validation as well as the test data set, and the metrics across the data sets is presented in Table \ref{table:performance metrics}. The table also illustrates that our model is not overfitting as the performance on validation and test data set are almost consistent. 

\begin{table}[ht]
  \centering
  \caption{Summary of performance metrics on Test and Test+Validation datasets.}
  \label{table:performance metrics}
  \begin{tabular}{l cc cc l} 
    \hline\hline
    & \multicolumn{2}{c}{\textbf{Test}} & \multicolumn{2}{c}{\textbf{Test+Validation}} & \\
    \cline{2-3}\cline{4-5}
    \textbf{Metric} & $\gamma=-0.1$ & $\gamma=2$ & $\gamma=-0.1$ & $\gamma=2$ & \textbf{Description} \\
    \hline
    Accuracy  & 0.85 & 0.82 & 0.85 & 0.83 & $\tfrac{\text{TP}+\text{TN}}{\text{Total}}$ \\
    Precision & 0.86 & 0.82 & 0.85 & 0.82 & $\tfrac{\text{TP}}{\text{TP}+\text{FP}}$ \\
    Recall    & 0.83 & 0.82 & 0.85 & 0.82 & $\tfrac{\text{TP}}{\text{TP}+\text{FN}}$ \\
    F1-Score  & 0.85 & 0.82 & 0.85 & 0.82 & Harmonic mean of Precision and Recall \\
    \hline
  \end{tabular}

  \vspace{0.5em}
  {\footnotesize\emph{Note.} TP = true positive; FP = false positive; TN = true negative; FN = false negative.}
\end{table}

\begin{figure}
    \centering
    \includegraphics[width=0.45\linewidth, height=7cm]{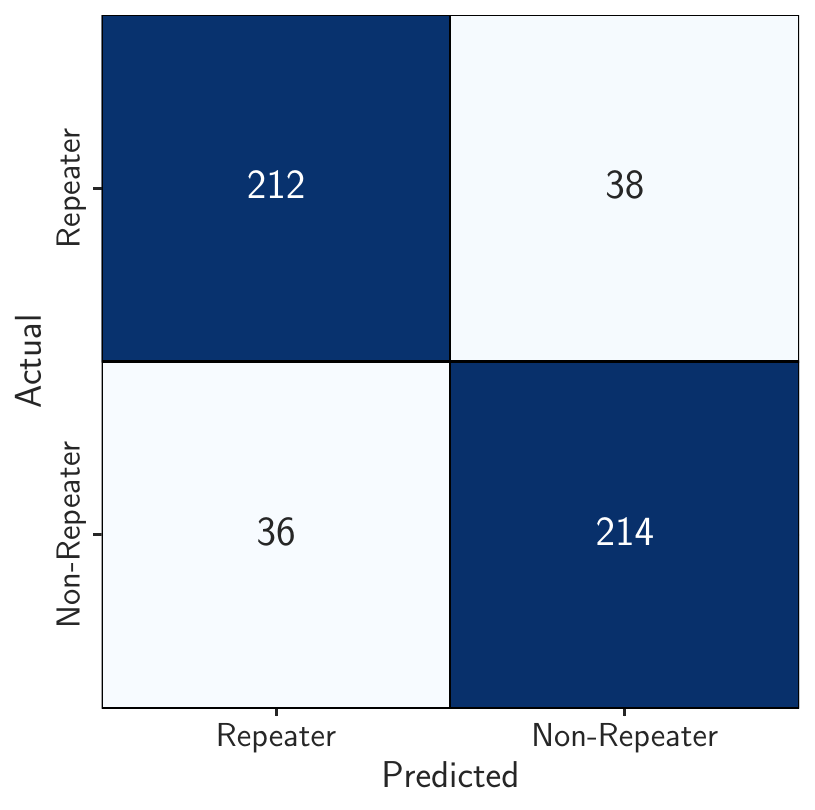}
    \hfil
    \includegraphics[width=0.45\linewidth, height=7cm]{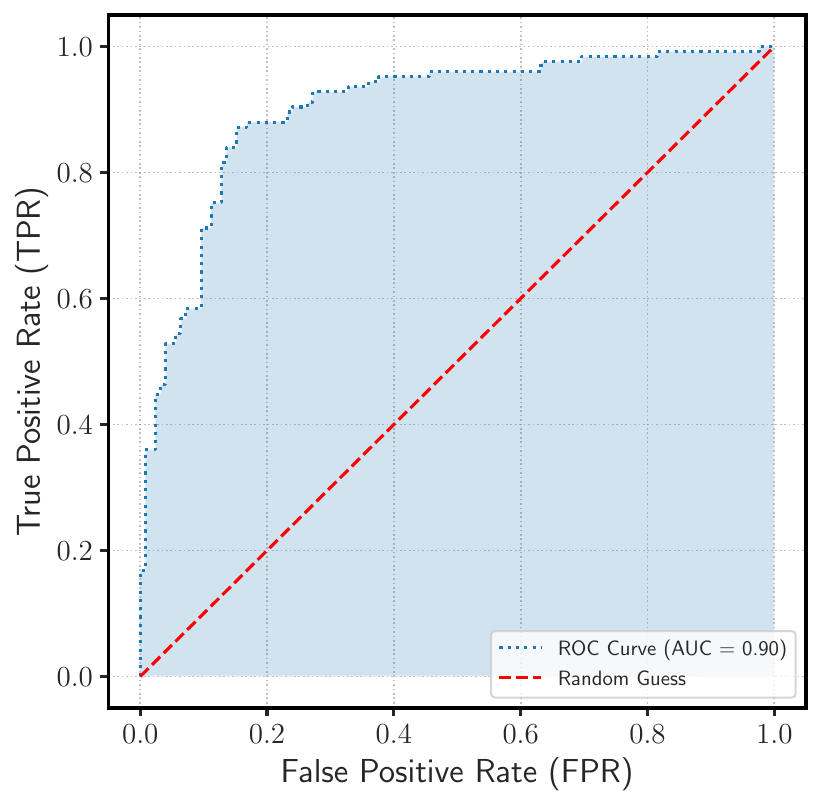}
    \caption{(Left) Confusion matrix for the test and validation data set. Test and Validation data set are combined as the performances of model on both of them are almost similar and consistent and (Right) receiver operating characteristic (ROC) curve on the test data set for $\gamma=-0.1$ model.}
    \label{fig:conf-matrix and roc}
\end{figure}

Clearly the model with $\gamma=-0.1$ outperforms the model with $\gamma=2$ in all the performance metrics. Furthermore, the model also outperforms in recall, i.e., its ability to truly predict a repeater as a repeater. The reason for improved performance when using $\gamma < 0$ might be due to inherent limitations in the data set, e.g., mislabeling due to lack of repeat bursts. Some of the events labeled as non-repeaters might be actually from a repeating source that has not yet emitted a second detectable FRB, and also some of the events associated with a repeating source might not be from the source due to the poor localization capabilities of CHIME \citep{Amiri_2018}. The harder to classify samples are those that are mislabeled \revieweredit{or those with transformed spectrum due to off-detection from center of formed beams \citep{CHIMEFRB_Collaboration2020a}}, and the model with the negative value of focusing parameter is slightly more focused on easy to classify samples. Due to the presence of harder to classify samples in our data set and higher performance metrics of model with negative value focusing parameter, it is straightforward to implement the model  with $\gamma=-0.1$ for inferences and analyses. A confusion matrix for $\gamma=-0.1$ on test and validation data set combined is shown in the Figure \ref{fig:conf-matrix and roc}  (Left). \revieweredit{The data set are combined as the performance metrics were consistent over both the data sets as presented in Table \ref{table:performance metrics}, illustrating model didn't overfit during the training.}

Figure \ref{fig:conf-matrix and roc} shows receiver operating characteristic (ROC) curve for the test data set and the area under the curve is 0.9, demonstrating that the ConvNext model with $\gamma = -0.1$ has a reasonable ability to distinguish between repeating FRBs and apparently non-repeating FRBs based solely on the morphology contained in the supplied images. As our deep learning CNN-based model was able to classify repeating and apparently non-repeating FRBs with good performance metrics, it suggests the existence of morphological variation in dynamic spectra between the two sub-subclasses of FRBs in \cattwo. This finding is consistent with the previous analyses \citep{Pleunis2021}  on \cfrb \catone. 

Our \convnext model with $\gamma=-0.1$ can be useful in making predictions about FRB events before making follow-up observations, which should give some heuristic probability values for future \cfrb observations. \referee{As a demonstration, there were $6$ repeating FRBs in the test and validation dataset which were labeled as non-repeaters in CHIME/FRB \catone. The deep learning model predicted $4$ of them as repeaters which are shown in Figure \ref{fig:nr_cat_1_r_cat_2_true_pred}.
\begin{figure}
    \centering
    \includegraphics[width=0.8\linewidth]{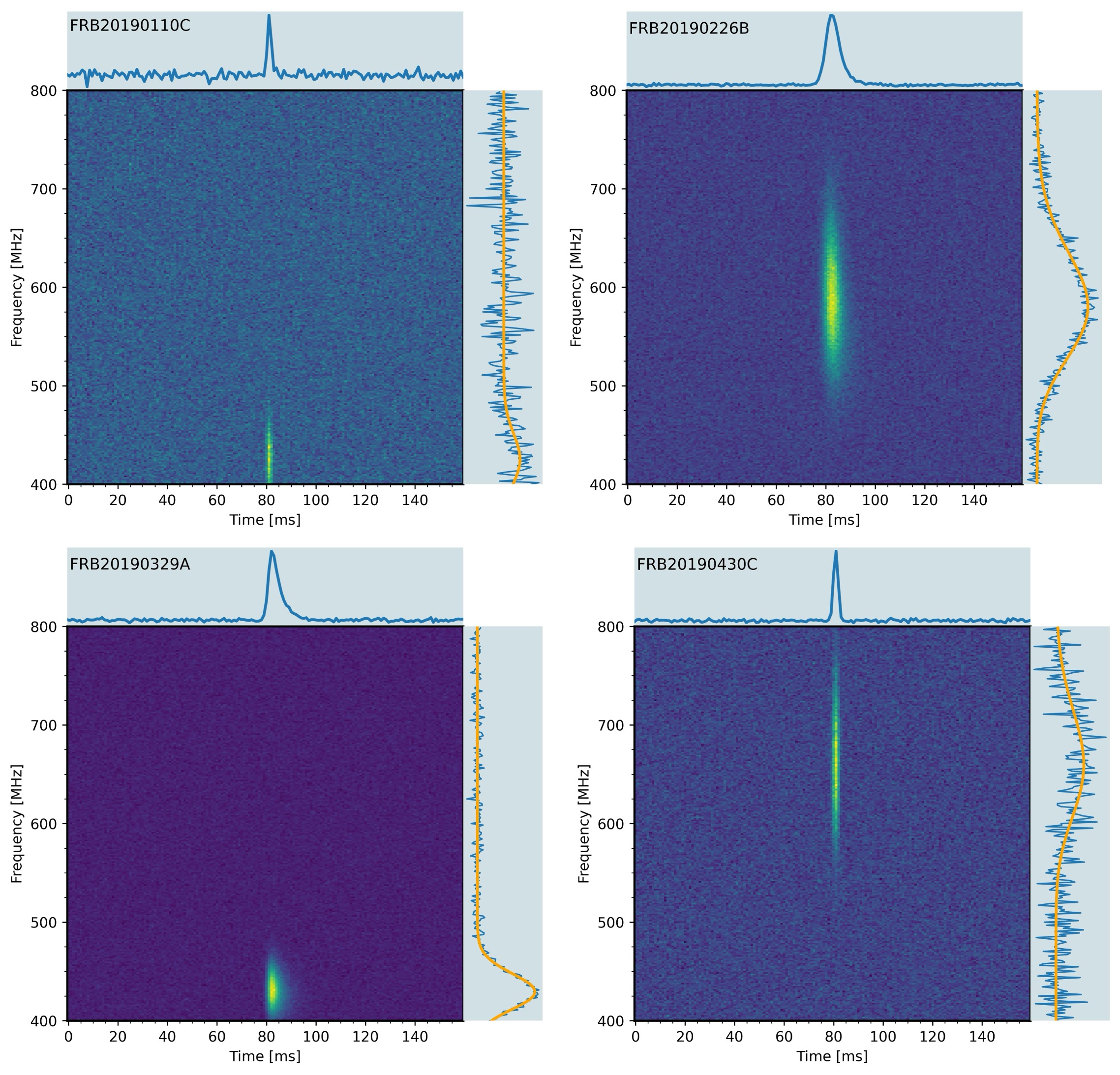}
    
    \caption{Non-repeating events in \cfrb \catone which are predicted to be repeater by the deep learning model are confirmed as such in \cfrb \cattwo.}
    \label{fig:nr_cat_1_r_cat_2_true_pred}
\end{figure}
Other bursts from these sources were observed and are labeled as repeaters in CHIME/FRB \cattwo.
\begin{figure}
\centering
\includegraphics[width=0.8\linewidth]{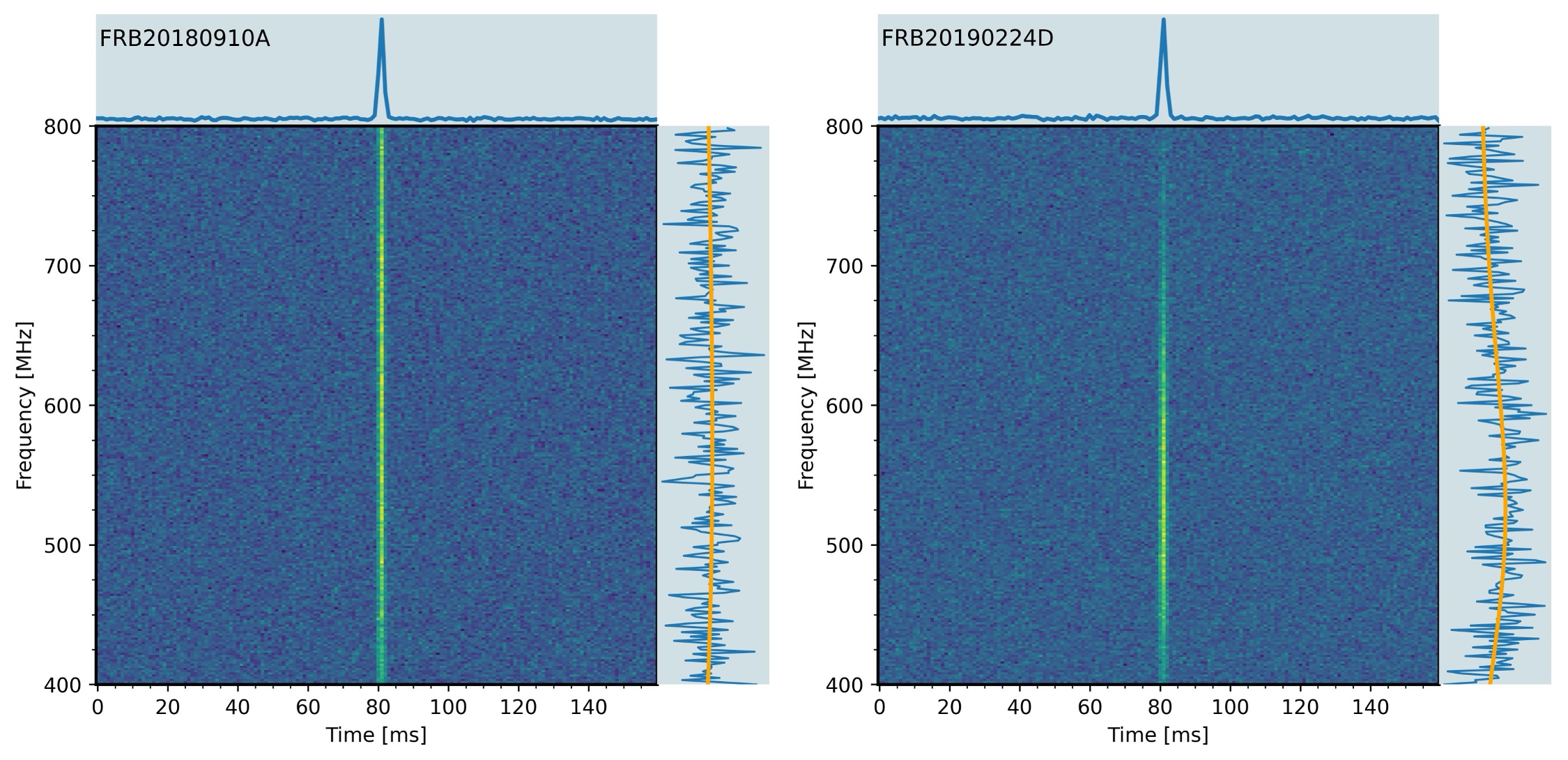}
\caption{Non-repeating events in \cfrb \catone which are predicted to be non-repeater by the deep learning model are actually repeater in \cfrb \cattwo.}
\label{fig: cat1_nr_cat2_r_false_pred}
\end{figure}
The remaining $2$ were predicted to be non-repeating  which are shown in Figure \ref{fig: cat1_nr_cat2_r_false_pred}. Both of these events are broad-band in frequency and narrow in time suggesting that some repeating source starts off with a burst  resembling to that of non-repeating in morphology. This advocates for morphology evolution study of repeating FRBs using deep learning tools.}

\begin{figure}
    \centering
    \includegraphics[width=0.8\linewidth]{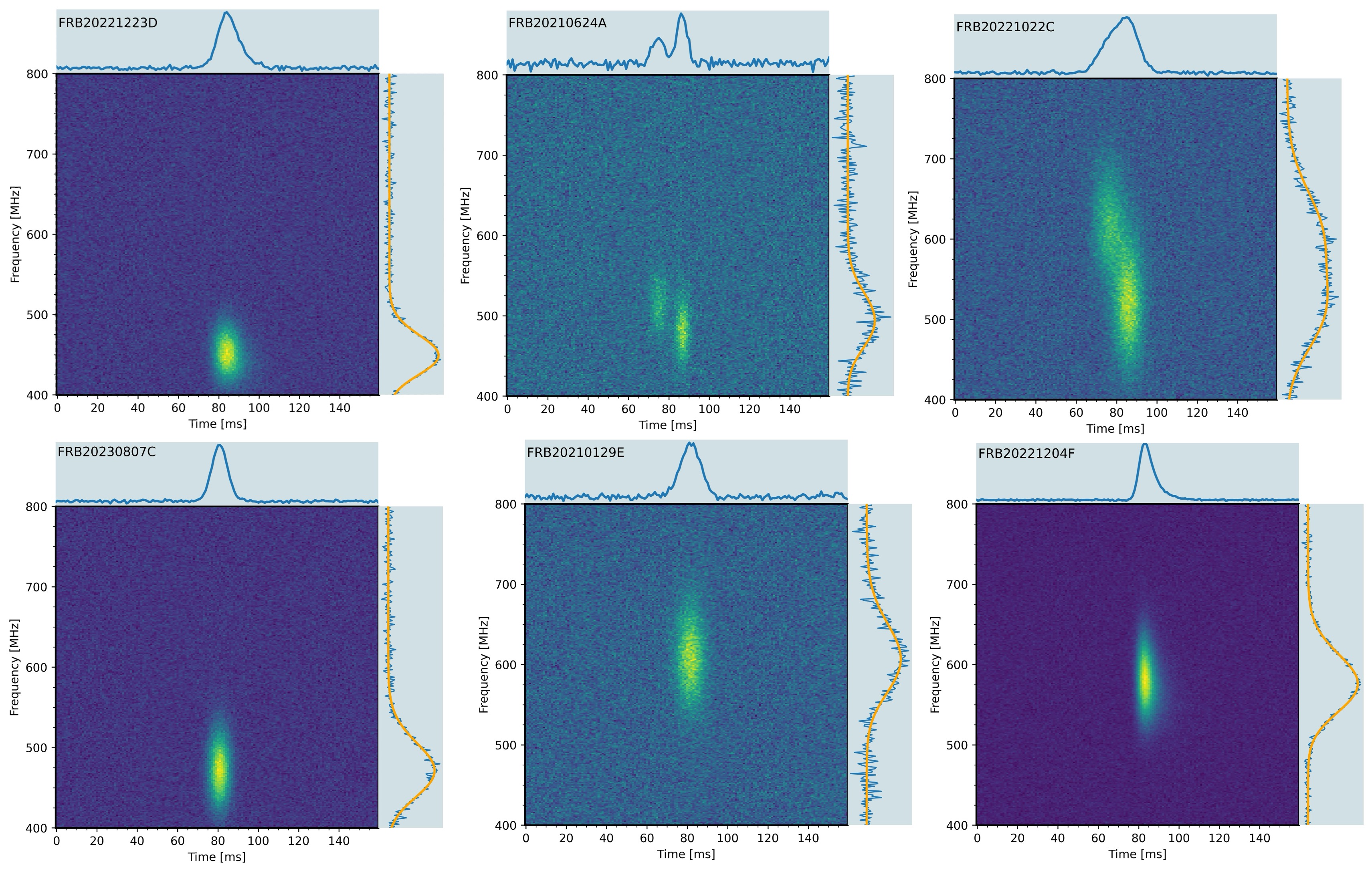}
    
    \caption{Repeaters correctly classified as repeater.}
    \label{fig: r as r}
\end{figure}
\begin{figure}
    \centering
       \includegraphics[width=0.8\linewidth]{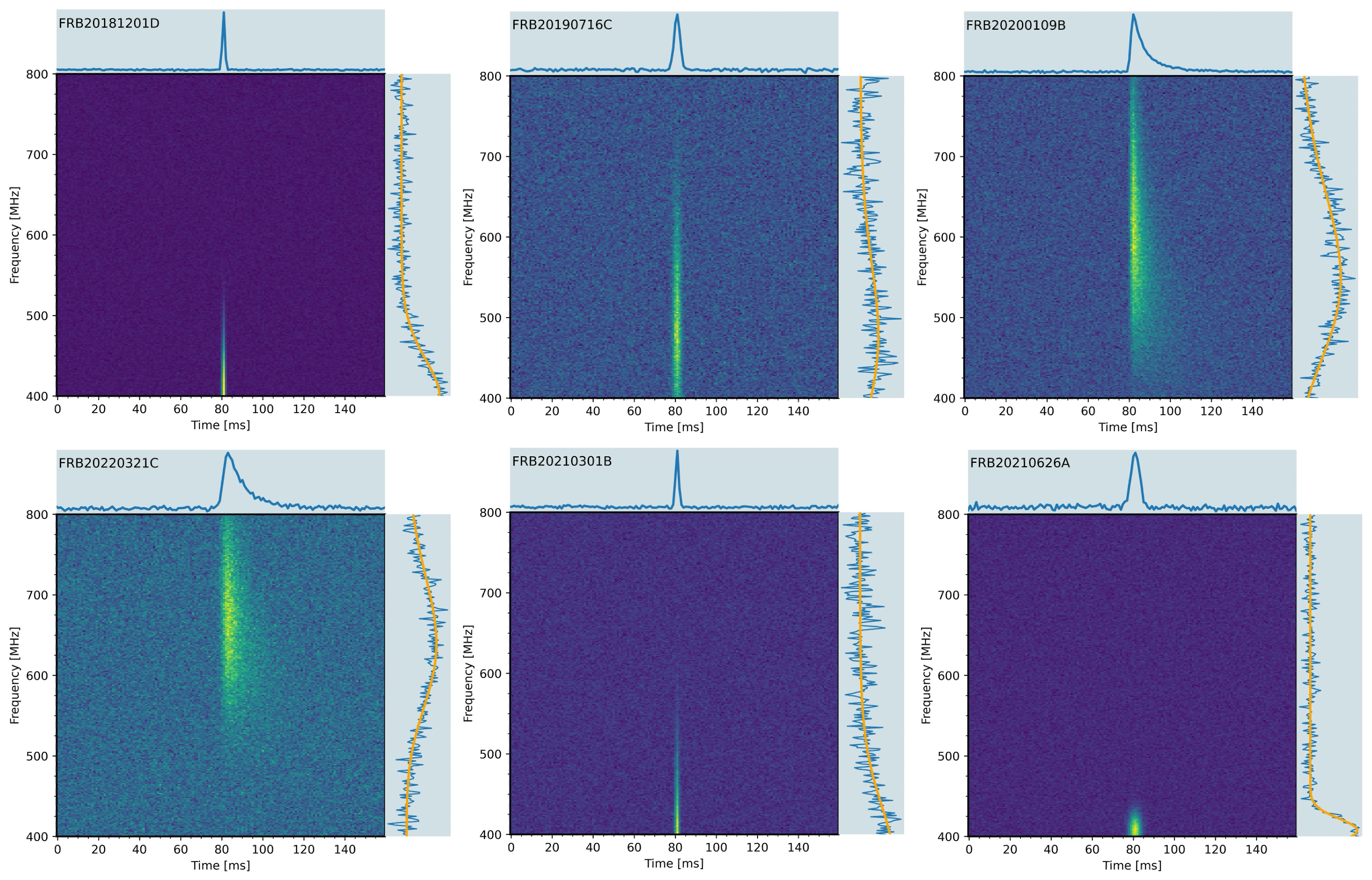}
    \caption{Repeaters incorrectly classified as non-repeaters}
    \label{fig: r as nr}
\end{figure}

Figure \ref{fig: r as r} shows examples of known repeaters from the test data set that were predicted correctly by the best model (model with $\gamma=-0.1)$. These events consist of bursts that are narrow banded in frequency and broader in time. 
Figure \ref{fig: r as nr} shows FRBs from known repeating sources that are incorrectly predicted as non-repeating. These events are  mostly band-edged, narrow in time and broad-band, and/or faint. 

\begin{figure}
    \centering
    \includegraphics[width=0.8\linewidth]{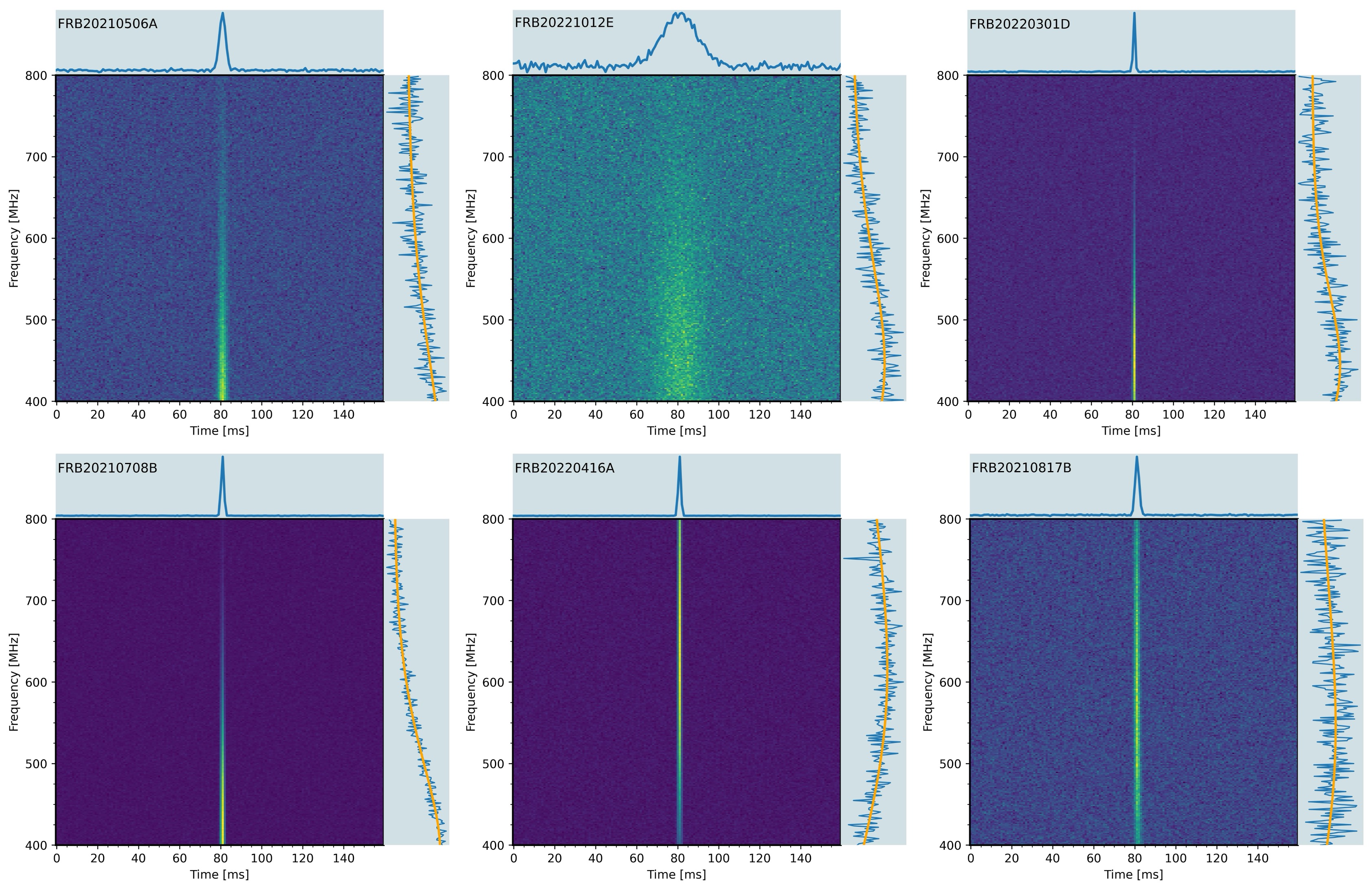}
    \caption{Non-repeaters correctly classified as non-repeaters.}
    \label{fig:nr as nr}
\end{figure}
\begin{figure}
    \centering
    \includegraphics[width=0.8\linewidth]{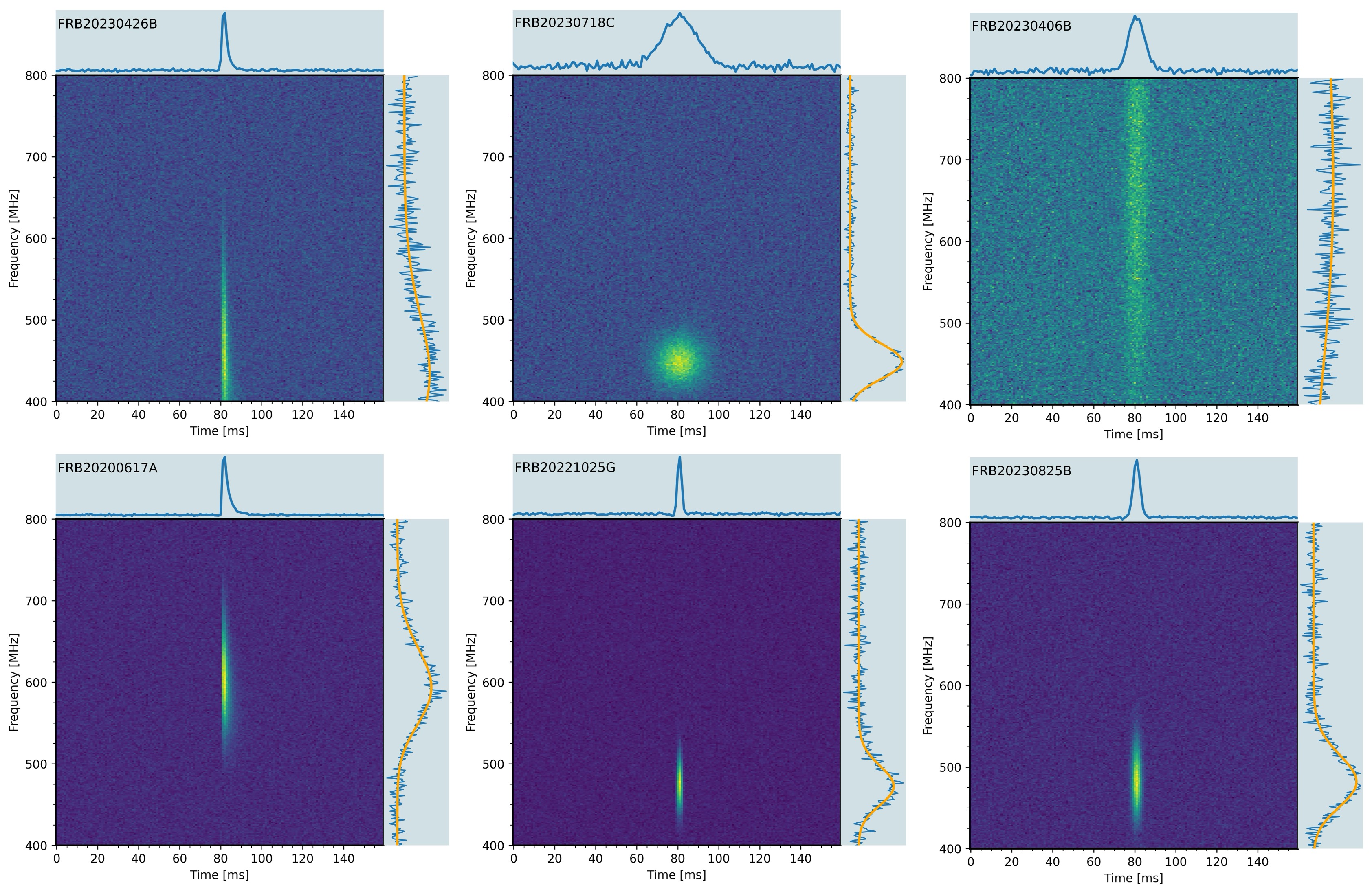}
    \caption{Non-repeaters incorrectly classified as repeaters.}
    \label{fig:nr as r}
\end{figure}

Similarly, Figure \ref{fig:nr as nr} displays events from the test data set that were correctly classified as non-repeaters. These events are broad-band in frequency and narrow in time. Figure \ref{fig:nr as r} are the events misclassified as repeaters. Some of these events are faint, some are narrow in frequency, and some resemble the features of repeating FRBs, indicating potential repeating candidates. 

\subsection[Comparison with parameter based simple machine learning algorithms]{
    \referee{Comparison with parameter based simple machine learning algorithms}
}
\referee{We also performed parameter based classifications using simple supervised machine learning algorithms to compare against the results obtained by using deep learning. The parameters that were used are spectral index, spectral running, burst width, boxcar width, lower frequency, upper frequency and peak frequency. Other non-morphological features like dispersion measure and signal to noise features like flux and fluence were not included in the analysis.

The performances of the various machine learning algorithms on the test dataset are presented in Table \ref{tab:mlalgorithms}.
Clearly, our deep learning model achieved an F-1 score  that exceeds the scores of all parameter based supervised methods listed in Tabe \ref{tab:mlalgorithms} by at least $10\%$. This demonstrates the necessity of a powerful feature extractor which doesn't rely on handcrafted morphological parameters but instead extracts subtle features from the dynamic spectrum which can be used for classification and other downstream tasks. Most of the deep learning feature extractors require a huge number of training dataset, however one can employ transfer learning requiring less data and little computing to achieve decent performance metrics as we have illustrated in this work.}

\begin{deluxetable}{l c c c}
\tablecaption{Performance metrics with different machine learning approaches}
\tabletypesize{\footnotesize}
\tablewidth{0pt}
\tablehead{
  \colhead{Algorithm} & \colhead{Precision} & \colhead{Recall} & \colhead{F-1 Score}
}
\startdata
Logistic Regression & 0.64 & 0.84 & 0.72 \\
K-Nearest Neighbors & 0.75 & 0.70 & 0.73\\
Random Forest & 0.75 & 0.70 & 0.72  \\
Gradient Boosted Decision Trees & 0.79 & 0.71 & 0.75 \\
Support Vector Machine & 0.54 & 0.82 & 0.65 \\
Linear Discriminant Analysis & 0.72  & 0.52 & 0.60  \\
\enddata
\tablecomments{To overcome the issue of unbalanced classes, ``balanced"  class weight was used when training the algorithms wherever it was applicable.}
\label{tab:mlalgorithms}
\end{deluxetable}

\subsection{Model Interpretation with Integrated Gradients}
\label{subsec:model_interp_real} 
The black-box nature of deep-learning models can obfuscate the factors that impact decision-making in classification. It is thus crucial to ensure that our model is accurately identifying important features, and is not being influenced by irrelevant noise elements. \revieweredit{One method to address this is through the use of integrated gradients as explained in Appendix \ref{appendix:integrated gradient}, which can help attribute the contribution of each feature to the model's prediction.}

\begin{figure}[h!]
\centering
\begin{tabular}[t]{@{}c@{\hspace{0.05\linewidth}}c@{}}
\begin{minipage}[t]{0.4\linewidth}
  \vspace{0pt} 
  \includegraphics[height=6.2cm,width=\linewidth]{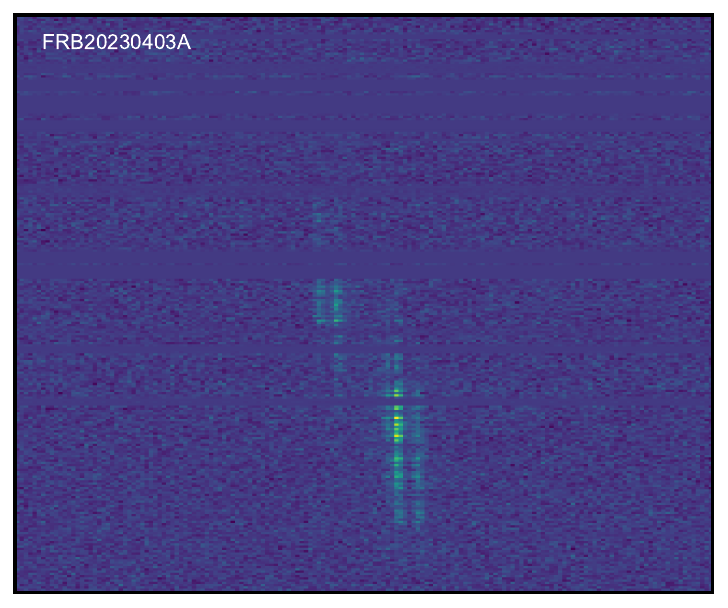}
\end{minipage} &
\begin{minipage}[t]{0.4\linewidth}
  \vspace{0pt} 
  \includegraphics[height=7cm,width=\linewidth]{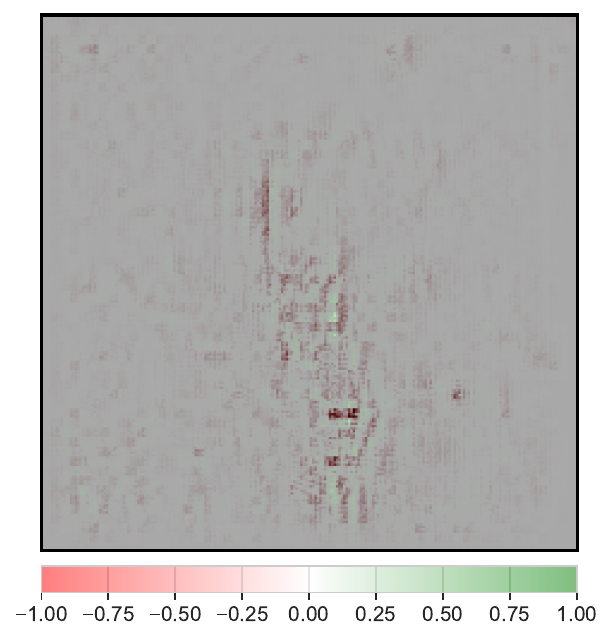}
\end{minipage} \\
\small{(a)} & \small{(b)}
\end{tabular}
\caption{(a) A repeater sample from the test data set on which the model made the correct prediction (b) Overlaid integrated gradient with the color bar representing the attribution magnitude.}
\label{fig:integrated gradients}
\end{figure}

 Figure \ref{fig:integrated gradients}(a) is the original input image from the test data set that was correctly  predicted as  a repeater with high confidence by the model. Figure \ref{fig:integrated gradients}(b) is the corresponding overlaid integrated gradient. The green and red pixels on the overlaid integrated gradients highlight the most significant features of the input image, which is used in making a decision by the model. On comparing the input image and overlaid integrated gradient, we can observe that the most important features are the pixels that contain bursts in the image. This \revieweredit{builds confidence} that the model is focusing on the correct areas for making predictions and decisions.

\section{Fine Tuning on Fitburst Models} \label{sec:fine tuning on fitburst modeled data}
The performance of machine learning models relies heavily on the representation of data on which they are trained \citep{bengio2014representationlearningreviewnew}. In the present context, ``representation" refers to the processing of raw data to minimize noise elements while retaining the valuable information in the raw data. The \convnext model developed in Section \ref{sec:CNN Model For chime-frb images} contains blocks with nonlinear transformations that enact feature (i.e., representation) extraction of the data; this model adequately classifies FRBs based on morphology as shown in Section \ref{sec: training_real}. However, we nonetheless sought to test if the model still improves upon feeding it with some preprocessed (i.e., represented) data.

\begin{figure}[h!]
    \centering
    \includegraphics[width=0.75\linewidth, keepaspectratio]{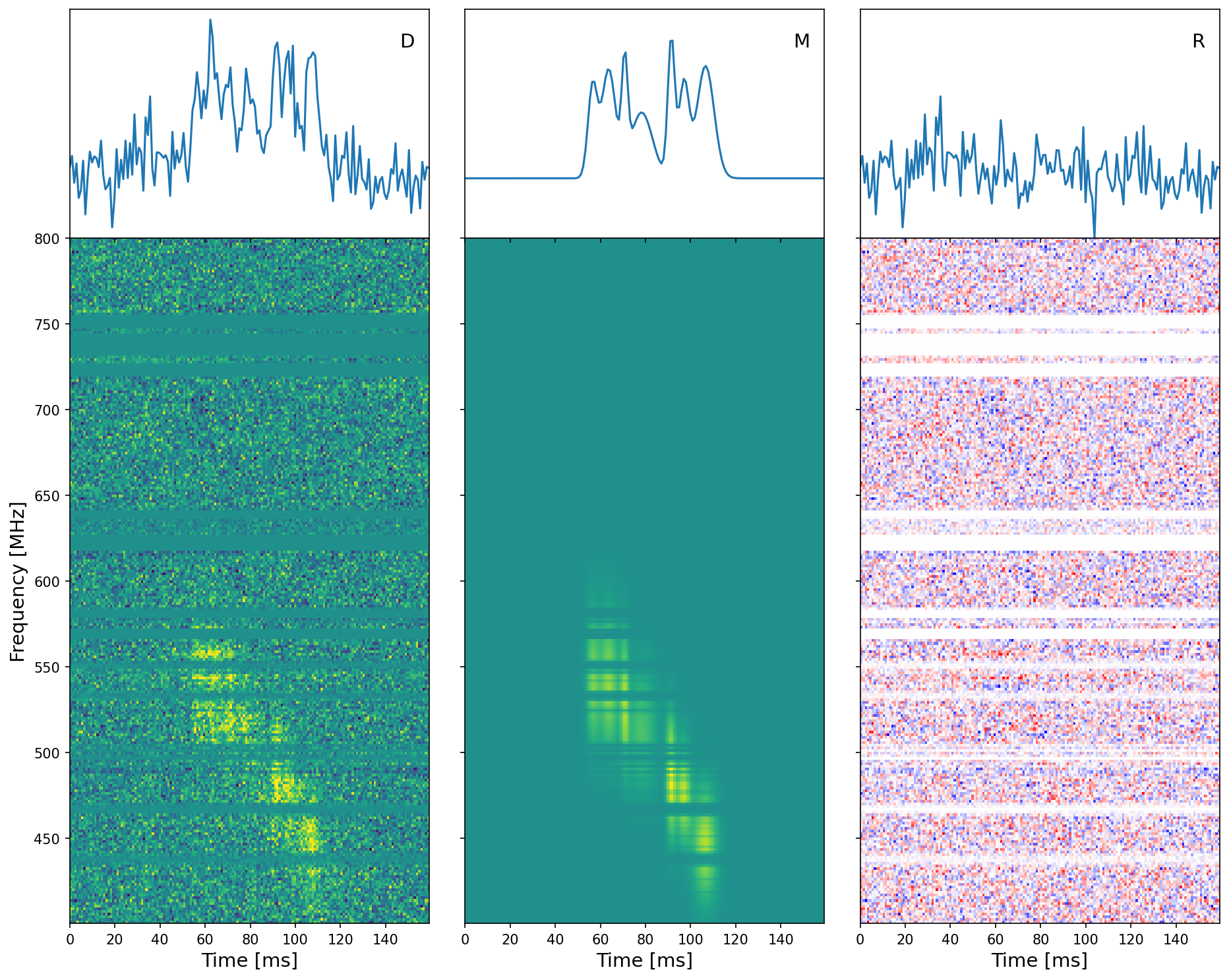}
    \caption{(Left) Dynamic spectrum of FRB20210523C. (Middle) synthetic representation of the dynamic spectrum modeled by \texttt{fitburst}. (Right) residual. }
    \label{fig:fitburst model}
\end{figure}

There are several approaches for obtaining noiseless representations of the data. A common way involves the construction of an encoder-decoder network and then training it using unsupervised algorithms \citep{CHARTE202093}. The output of the encoder block gives some form of representation of the data, which can be used for other tasks like supervised classification, image reconstruction, noise reduction and so on. However, this approach is severely limited by computational cost and the lack of physical interpretation of the represented data. Due to such reasons, we used \texttt{fitburst} to create images that, based on a physically motivated parametric model, are representations of the total intensity data. In addition, the trained neural network model can be interpreted by training on a data set that has been created using some mathematical model. An example of a represented physical model for FRB20210523C of the total intensity data with \texttt{fitburst} is shown in Figure \ref{fig:fitburst model}. The first panel shows total intensity data, the second panel is the representation of the total intensity data, and the last panel is the residual representing the goodness of fit verified using a weighted-$\chi^2$ statistic. 
 
\subsection{Data Preparation and Preprocessing}
\label{subsec:modeled data preparation and preprocessing}
Synthetic representations of all \cfrb \cattwo events were generated using \fitburst fitted parameters. All the image preprocessing steps were similar as mentioned in Section \ref{subsec:real_data_preparation}. We again used the pretrained \convnext model as discussed in Appendix \ref{appendix:convnext architecture}, and implemented transfer learning as in Section \ref{sec: training_real}.
Our unbalanced noiseless synthetic data encountered overfitting issues even with weighted random sampling, due to a large number of weights and biases in the \convnext model and simple represented data.  To fix the overfitting issue, we randomly sampled 600 repeating and 600 one-off FRB events for training, 100 for testing and 100 for validation from each sub class, and repeated that for five times, in a manner similar to a bootstrapping procedure. We limited the bootstrapping procedure to 5 iterations due to substantial computational cost associated with fine tuning of a deep neural network. We therefore used five randomly-sampled data sets to fine tune the \convnext model and to avoid overfitting. Each sampled data set is denoted as \textit{sampling-1} through \textit{sampling-5}.

\subsection{Fine Tuning and Results}
\label{subsec:training modeled data}
The hyper-parameters used to fine tune the model on the synthetic data set were similar to those in Section \ref{sec: training_real}. We used $\gamma=-0.1$ for the focal loss function. 
\begin{figure}[h!]
    \centering
    \includegraphics[width=0.6\linewidth]{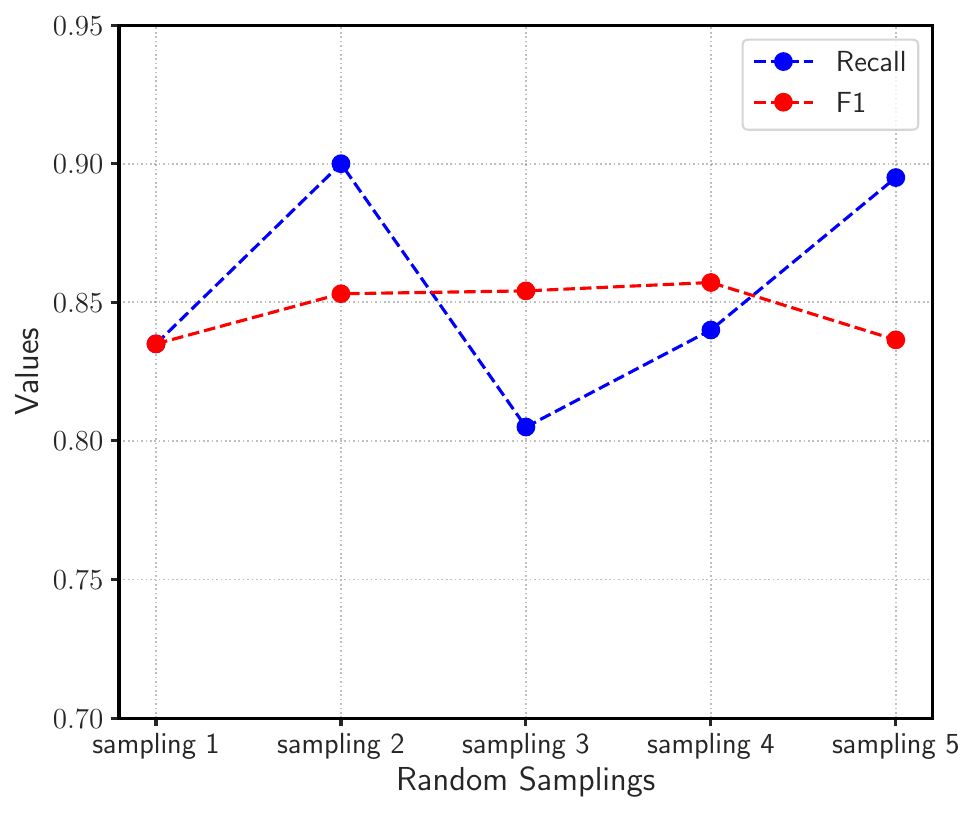}
    \caption{Performance metrics for test and validation data combined for each randomly sampled data set. Training and validation data set are combined as the evaluation performance metrics were similar on both.}
    \label{fig:f1_recall_model}
\end{figure}
The fine-tuning results evaluated for each of the randomly sampled data sets in the test and validation data are displayed in Figure \ref{fig:f1_recall_model}. All performance metrics are between 0.8 and 0.9, consistent with the \convnext model trained on total intensity images in Section \ref{sec: training_real}. From Figure \ref{fig:f1_recall_model}, the model fine tuned on \textit{sampling-2} data set outperforms models fine tuned on other samples of the data set on combined F1 and recall score. The difference in performance matrices on different samples suggests variability in the data set along with potential noise or outliers. Even though the model fine tuned on represented data shows no significant improvement over the model fine tuned on total intensity data, we can use the model trained on the represented data for interpretation. The model fine tuned on sampling-2 data set is used for model interpretation below. The results are likely to remain similar if any of the fine tuned models were used, as the performance metrics are not significantly different. 

\subsection{Model Interpretation}
Instead of using overlaid integrated gradients as before in Section \ref{subsec:model_interp_real} for model interpretation, we used \texttt{fitburst} to create synthetic representation of FRBs with \revieweredit{different physical properties by uniformly sampling across two burst parameters (while keeping the others constant). The parameters under consideration are burst width, spectral index, spectral running, number of sub-bursts in multi-component bursts, and sub-burst temporal separation. For the burst characteristics that were explicitly parametrized, uniform sampling was performed within the parameter ranges reported in \cfrb \cattwo.} Such synthetic bursts were passed through the \convnext model fine tuned on \fitburst representations of \cattwo FRBs, to see if any morphological properties were constrained by the model to associate a burst to one of the sub-class. The prediction probability of the model for different pairs of parameters are discussed below. However, we note that the values from the output of a deep neural network are only the heuristic notions of probability which represent the model confidence  rather than actual probability values. The threshold was set to 0.5, with values $\ge$ 0.5 corresponded to repeaters.

\subsection{Burst width vs Spectral running}
\label{subsec:bw vs sr}
Burst width is interpreted by \fitburst as the intrinsic width ($\sigma$) of a Gaussian pulse profile and spectral running ($\beta$) is a parameter in the running power law (RPL) model \citep{Fonseca2024} for frequency dependent spectral energy distribution (SED), given as
\begin{equation}
    F_k = \left(\frac{\nu_k}{\nu_r}\right)^{\phi + \beta\ln(\nu_k/\nu_r)}.
\label{eq:rpl}
\end{equation}

In Equation \eqref{eq:rpl}, $\nu_k$ is the $k$\textsuperscript{th} frequency channel, $\nu_r$ is the reference frequency and $\phi$ is the spectral index. As discussed by \cite{Fonseca2024}, repeater-like morphologies yield values of $\beta$ that serve as a measure of Gaussianity, and of $\phi$ that reflect offset between $\nu_r$ and the frequency of peak emission. Therefore, $\beta$ controls the spectral width and $\phi$ controls the position of the spectrum in the RPL model for frequency dependent SEDs. 

We used $\nu_r$ value of 600 MHz and $\phi$ value of 0 to create a synthetic representation for model interpretation, placing the peak emission at 600 MHz. If the value of parameter $|{\beta}|$ increases, the spectrum becomes narrower across frequency.

\begin{figure}[ht!]
    \centering
    \includegraphics[width=0.6\linewidth]{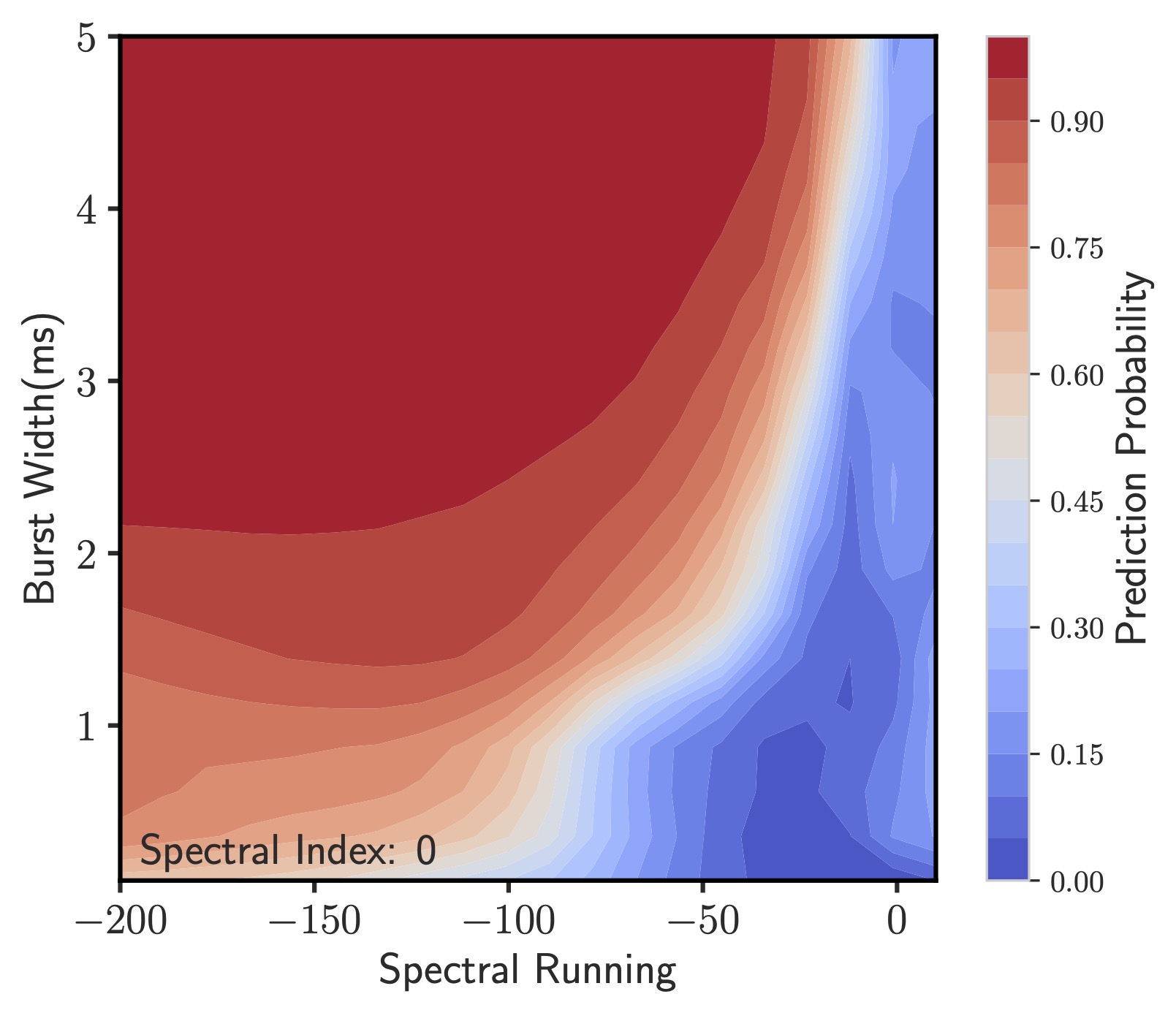}
    \caption{Model interpretation for different values of burst width and spectral running. The color scale represents prediction probability corresponding to each point in morphological phase space defined by spectral running and burst width with decision threshold of 0.5. The redder regions indicates a higher probability of being classified as repeater while bluer regions indicating a higher probability of being classified as non-repeater.}
    \label{fig:bw_vs_sr}
\end{figure}
Figure \ref{fig:bw_vs_sr} presents the model predictions for different values of burst width and spectral running. For $|\beta|>10$, the model predicts the classification of an FRB as a non-repeater with any value of width between 0 to 5 ms. As the spectrum becomes narrower (i.e., $|\beta| > 10$) and the width becomes larger, the \convnext model starts predicting a burst as a repeater. Therefore, a burst with narrow spectrum and wider temporal width is more likely to be classified as a repeater by the \convnext model. This effect has already been observed through statistical analysis of \cfrb \catone data \citep{Pleunis2021}. We note that values of $\beta$ in this work do not directly correspond to the spectral running values in the published \cattwo data set because the $\nu_r$ and $\phi$ values here are inconsistent with the values in \cattwo. This mismatch is not significant, because the reference frequency $\nu_r$ is an arbitrary value, leading to an overall offset in the distributions of spectral index and spectral running values. 

\subsection{Sub-burst separation vs Spectral running}
In this analysis, we varied temporal separation between sub-bursts and $\beta$ i.e. spectral width for two components. The spectral index and burst width for each components were held fixed to values $\phi=0$ and $\sigma=1$ms. 

\begin{figure}[ht!]
    \centering
    \includegraphics[width=0.6\linewidth]{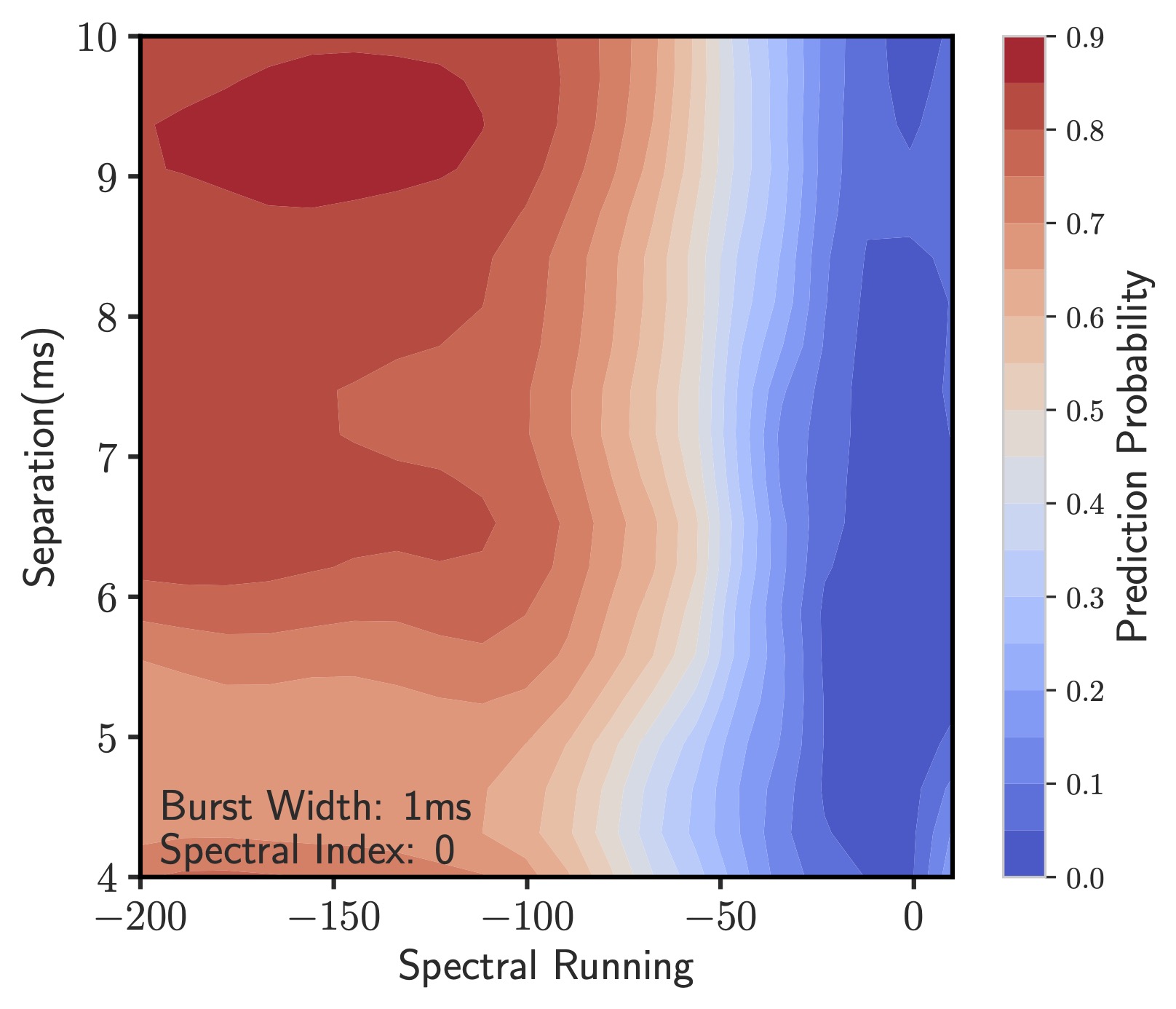}
    \caption{Interpretation of our \fitburst-trained \convnext model for different values of inter-component separation and spectral running for two component burst. The color scale represents prediction probability corresponding to each point in morphological phase space defined by spectral running and sub-burst separation with decision threshold of 0.5. The redder regions indicates a higher probability of being classified as repeater while bluer regions indicating a higher probability of being classified as non-repeater. This prediction map demonstrates that narrow frequency bursts with larger sub-burst separations are likely to be repeaters.}
    \label{fig:seperation_vs_sr}
\end{figure}
Figure \ref{fig:seperation_vs_sr} illustrates that the fitburst-trained \convnext model is insensitive to sub-burst separation for very narrow or very broad spectra. For some bandwidths ($|\beta|$ $\sim$ 60 - 80), the events are predicted to be non-repeaters for small separation but are predicted to be repeaters when the separation between components increases.

\subsection{Sub-burst drift vs Spectral running}
\label{subsec:drift vs spectral running}
Repeating FRBs with multiple components often exhibit frequency drifts \citep[e.g.,][]{Hessels2019}. In principle, we can use our fitburst-trained \convnext model to check how the trained model interprets bursts with multiple components that also display drifts. To simulate drifts in two component bursts, we made use of the parameter $\phi$ in Equation \eqref{eq:rpl}. Two components with different values of $\phi$ but other parameters kept the same will drift in frequency.

\begin{figure}[ht!]
    \centering
    \includegraphics[width=0.6\linewidth]{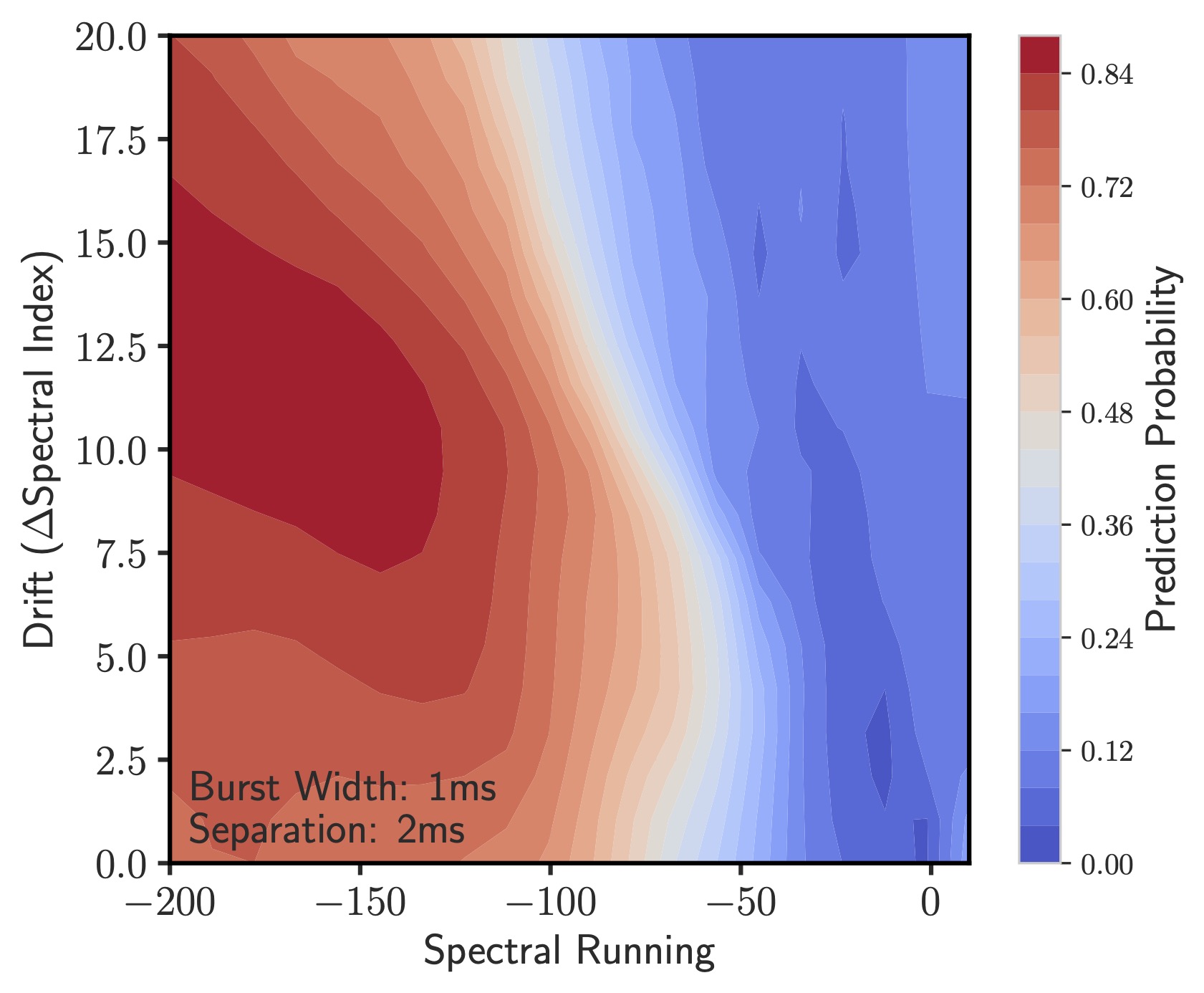}
    \caption{Model interpretation for different drift rate and spectral running. The color scale represents prediction probability corresponding to each point in morphological phase space defined by drift rate parametrized by separation in spectral index values between sub-bursts and spectral running with decision threshold of 0.5. The redder regions indicates a higher probability of being classified as repeater while bluer regions indicating a higher probability of being classified as non-repeater.}
    \label{fig:drift vs sr}
\end{figure}

Two component bursts with burst widths of 1 ms and separation of 2 ms were drifted by changing the value of $\phi$ for the second component. The larger the difference between $\phi$ of the two components, the larger the drift rate. Figure \ref{fig:drift vs sr} represents that both repeaters and non-repeaters are likely to display drifts. However, for some $\beta$ values in the range  $-90 < \beta < -60 $, the two-component bursts were predicted non-repeaters for no drift; and as the drift increased between components, bursts with the mentioned $\beta$ range were labeled as repeaters until the extent of drifting became large enough that bursts with steep drifting were again labeled as non-repeaters. This feature indicates that the \convnext model is interpreting multi-component bursts with little to medium drift as repeaters and very high drift as non-repeaters. \revieweredit{It should be noted that, the drift rate was not parametrized in \cfrb \cattwo. During our synthetic data generation, we introduced this parametrization and in doing so we may have generated the synthetic FRBs with higher drift rates than maximum drift rate in \cfrb \cattwo events. Consequently, our model predictions at higher drift rates may represent extrapolations beyond the actual observed parameter space. It could therefore be a future work to explicitly parametrize the drift rate and conduct a more rigorous analysis to validate our findings.}

\subsection{Sub-burst periodicity vs Spectral running}
The number of multi-component FRB events in the \cfrb \cattwo has increased significantly from \catone \citep{chimefrb2021}. It becomes quite natural and essential to test the sub-burst periodicity of bursts with more than two components and their attribution to a sub-class of FRB. Here, we only interpret our \convnext model on three component bursts. If $t_1$ is the arrival time of the first component and $t_3$ is the arrival time of the third component, we can parametrize the arrival time of the second component $t_2$ in terms of a periodicity parameter $\theta$ as
\begin{equation}
    t_2 = (t_1 + \alpha) + \left(\frac{1 + \sin\theta}{2}\right)\times(t_3-t_1+2\alpha);\quad \frac{\pi}{2} \le \theta \le \frac{3\pi}{2},
\label{ed:periodicity}
\end{equation}
where $\alpha$ in \eqref{ed:periodicity} is an offset parameter that represents the minimum separation between the second component and all others. The value $\theta=\pi$ in Equation \eqref{ed:periodicity} places the second burst exactly in between $t_1$ and $t_3$, creating periodic sub-bursts. 
\begin{figure}[ht!]
    \centering
    \includegraphics[width=0.6\linewidth]{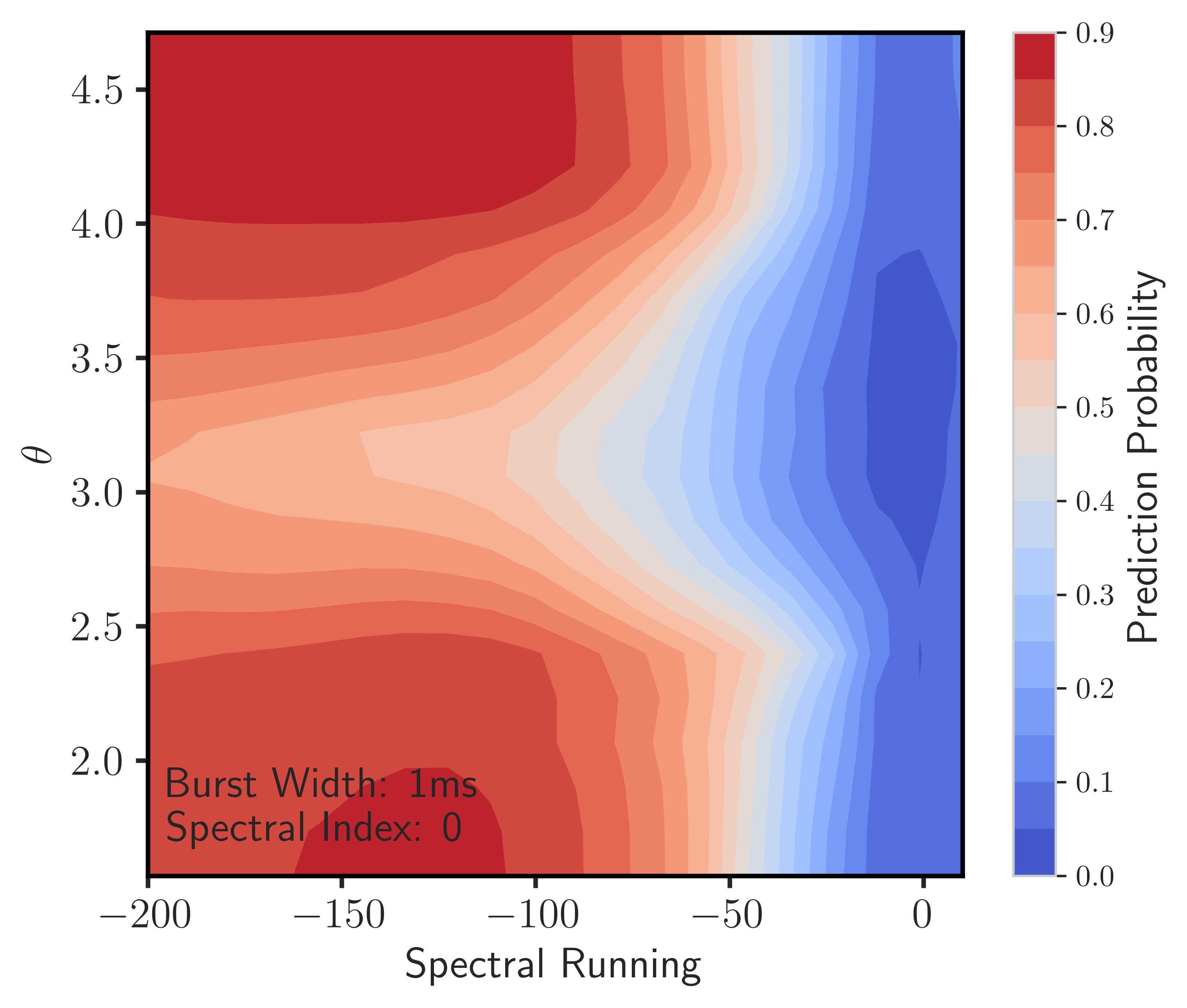}
    \caption{Model interpretation for different values of sub-burst periodicity parameter and spectral running. The color scale represents prediction probability corresponding to each point in morphological phase space defined by periodicity parameter and spectral running with decision threshold of 0.5. The redder regions indicates a higher probability of being classified as repeater while bluer regions indicating a higher probability of being classified as non-repeater.}
    \label{fig:periodicity vs sr}
\end{figure}
Clearly from Figure \ref{fig:periodicity vs sr}, the \fitburst-trained \convnext model is suggesting tentative periodicity in sub-bursts for non-repeating FRBs. As $\theta$ value move away from $\pi$, the model becomes increasingly confident that an event is a repeater. \revieweredit{As in the case of sub-burst drift in Section \ref{subsec:drift vs spectral running}, the periodicity was not explicitly parametrized in \cfrb \cattwo. However, there were events with three components in the training data so the prediction based on periodicity is a reflection of data distribution itself rather than bias imposed by the CNN model. This is again a subject of more rigorous analysis and a potential future work.}

\section{Discussions and Conclusions}
\label{sec:discussions and conclusions}
The main goal of this study was to implement a deep learning approach to characterize the two sub-populations of \cfrb \cattwo events \revieweredit{based purely on morphological features. As described in Sections \ref{sec:CNN Model For chime-frb images}} and \ref{sec: training_real}, we first trained a CNN on raw, labeled FRB images and showed that such a model can successfully classify events with an accuracy \revieweredit{over 85\% based purely on morphology}. We were able to \revieweredit{achieve successful} classification of repeating and non-repeating FRBs using a transfer learning approach with only a few layers activated during training \revieweredit{given the amount of noise in the data, high class imbalance ratio ($\sim$ 1:5), and taking account of only morphological features.} \revieweredit{We then explored using \fitburst{} representations of \cattwo{} events when training the \convnext{} model, in order to enable model interpretation.}


\subsection{\revieweredit{Implications from Model Performance}}

One of the major implications of our analysis is that \revieweredit{image-trained} models like those developed in Sections \ref{sec: training_real} and \ref{sec:fine tuning on fitburst modeled data} can be used to identify a repeater for follow-up observations\revieweredit{, or to associate} a new FRB event to a repeater with similar coordinates and DM. This analysis is significant as it proposes a novel\revieweredit{, data-driven} way to characterize FRBs based on morphology \revieweredit{and in a manner independent of morphological models. In this sense, we consider our \convnext models to be an improvement over} previous \revieweredit{machine-learning} analyses \revieweredit{that} used parameter-based algorithms\revieweredit{, as the use of model parameters likely ignores} complex burst morphologies and relationships between sub-pulses. 



\subsection{Impacts of Interpretability on Physical Implications}
The high performance of \revieweredit{our image-based} CNN models on classification suggests the existence of morphological differences between the repeating and apparently non-repeating FRBs. This finding \revieweredit{remains} consistent with previous studies \revieweredit{that used CHIME/FRB total intensity data} \citep[e.g.,][]{Pleunis2021, curtin2024morphology32repeatingfast} \revieweredit{even as the data set has grown by an order of magnitude. This consistency therefore means that past physical inferences have remained unchanged -- the sub-population of repeaters preferentially emit radiation over low bandwidths and therefore reflect an emission mechanism distinct from those of non-repeaters.}

\revieweredit{The use of \fitburst to generate synthetic bursts in deep learning allowed for model interpretation, as described in Section \ref{sec:fine tuning on fitburst modeled data}. This interpretation can be used to assess the correlation (or lack thereof) of morphological properties with classification likelihood. For example, when trained on \texttt{fitburst}-generated images} our \convnext model exhibited the possibility of sub-pulse drifts being associated with non-repeaters \revieweredit{while previous studies have suggested that downward drifts of sub-pulse were associated strictly with repeaters\citep{Wang_2019, Pastor_Marazuela_2023}.} 

\revieweredit{Our interpretable \convnext model also illustrated hints of quasi-periodicity at sub-second scales being associated with non-repeating FRBs, and non-periodicity being associated with repeating FRBs. While marginal in significance, this result would serve as an interesting avenue for exploring viable FRB progenitors should future data sets strengthen any potential correlation. We merely highlight that such questions can be addressed using CHIME/FRB images and our interpretable \convnext model.}

\subsection{Model Limitations}

Despite the \revieweredit{high-accuracy performance of our \convnext models}, this study should be \revieweredit{viewed} in light of \revieweredit{possible} limitations. \revieweredit{The major} limitation \revieweredit{in our work} is the lack of uncertainty on the inferences by the \convnext models. The \texttt{fitburst}-represented data were sub-sampled randomly for training\revieweredit{, in order} to overcome the problem of overfitting, which might have created some bias in the training data. \revieweredit{Furthermore,} only two degrees of freedom were used to generate synthetic bursts during the interpretation of the model trained on the representation data set\revieweredit{;} in reality\revieweredit{,} an FRB event and its association to a sub-class is a function of more than two physical parameters, which might again have biased our results. 

\revieweredit{Another significant limitation is the frequency-dependent nature of CHIME sensitivity as a function of sky location: the ``true" burst morphology may be different from what is observed in total-intensity data due to the frequency-dependent gain of both the CHIME primary and synthesized beams \citep{Pleunis2021}. This discrepancy arises especially when the bursts are detected off-axis, where detections can display ``spectral knots" when observed at far sidelobes \citep{lsc+24}. This bias can be resolved by using ``baseband" (i.e., raw voltage) data  and could be a potential future work to train a deep learning model on such data for better reliability and performance.}

\revieweredit{Finally, we also acknowledge the possibility of labeling error due to an inadequate timespan of our data for resolving FRBs from repeating sources. In particular, some fraction of ``true" repeating FRBs may not have emitted detectable repeat bursts and are therefore labeled as apparent non-repeaters in \cattwo. This possible contamination will affect the accuracy of our \convnext models. This diminishes with future catalogs but inevitably relies on an improved understanding of repetition rates, which is the subject of ongoing work.}

\subsection{\revieweredit{Prospects for Training on Non-morphological Features in FRB Images}}
\revieweredit{The models we developed in this work relied on de-dispersed, total-intensity spectra and do not rely on an interpretation on pseudo-morphological\footnote{In the context of our present work, we do not consider these parameters to be morphological as they fundamentally arise from plasma-astrophysical processes associated with intervening electronic media, and not the underlying FRB emission mechanism.} \fitburst{} parameters, i.e., DM, scattering timescale, and polarization ``rotation measure" (RM).} Training a machine learning model with \revieweredit{any of these} non-morphological parameters can increase the performance of classification metrics, but \revieweredit{results are likely to be} biased \revieweredit{as present data yield a} very small sample of repeating sources as compared to the sample size of non-repeating sources. 

\revieweredit{We nonetheless consider it worthwhile to gradually expand frameworks such as ours to incorporate these non-morphological properties into machine learning analyses on image data. For example, there is a small and growing population of ``persistent radio sources" \citep[PRSs;][]{lca22} that so far are only associated with repeaters \citep{mph+17,nal+22,bpy+24,bpy+25}. While clearly small as a sample, these repeaters also exhibit large magnitudes in RM\citep[e.g.,][]{gly25}; if this trend remains true for the broader sub-population, then the morphology in Stokes-polarization dynamic spectra can be used as another discriminant for classification via deep learning. Such analyses will require the use of voltage data that retain the information needed to derive polarization measurements, of which the CHIME telescope is producing a growing catalog for FRB measurements \citep{aaa+25}.}

\section*{acknowledgments}
We acknowledge that CHIME is located on the traditional, ancestral, and unceded territory of the Syilx/Okanagan people. We thank Bryan Gaensler for useful comments on this work.

We are grateful to the staff of the Dominion Radio Astrophysical Observatory, which is operated by the National Research Council of Canada.  CHIME is funded by a grant from the Canada Foundation for Innovation (CFI) 2012 Leading Edge Fund (Project 31170) and by contributions from the provinces of British Columbia, Qu\'ebec and Ontario. The CHIME/FRB project is funded by a grant from the CFI 2015 Innovation Fund (Project 33213) and by contributions from the provinces of British Columbia and Qu\'ebec, and by the Dunlap Institute for Astronomy and Astrophysics at the University of Toronto. Additional support is provided by the Canadian Institute for Advanced Research (CIFAR), McGill University and the McGill Space Institute thanks to the Trottier Family Foundation, and the University of British Columbia. The CHIME/Pulsar instrument hardware is funded by the Natural Sciences and Engineering Research Council (NSERC) Research Tools and Instruments (RTI-1) grant EQPEQ 458893-2014.

E.F. is supported by the National Science Foundation under grant AST-2407399. P.S. acknowledges the support of an NSERC Discovery Grant (RGPIN-2024-06266). D.C.S. is supported by an NSERC Discovery Grant (RGPIN-2021-03985) and by a Canadian Statistical Sciences Institute (CANSSI) Collaborative Research Team Grant.

\software{
Numpy \citep{harris2020array},
Matplotlib \citep{Hunter:2007},
PyTorch \citep{paszke2019pytorch},
torchvision \citep{torchvision2016},
pandas \citep{reback2020pandas},
seaborn \citep{Waskom2021}, 
Captum \citep{kokhlikyan2020captum},
\fitburst  \citep{Fonseca2024}
}

\revieweredit{
\appendix
\section{Convolutional Neural Networks}
\label{appendix:cnn}
The design of a CNN draws inspiration from the way the animal visual cortex processes visual data. Initially developed in the 1990s for handwritten digit classification \citep{lecun1990handwritten}, CNNs have proven to be highly effective for tasks such as image recognition, object detection, and video analysis. CNNs are typically made up of multiple convolution layers, with each layer responsible for detecting distinct features. For a 2-dimensional input image X of size $n_1 \times n_2$ and a kernel W of size $k_1 \times k_2$, the discrete convolution $Y = X * W$ between X and W is defined mathematically as:
\begin{equation}
Y(i, j) = X * W  =  \sum_{k_1} \sum_{k_2} X(i-k_1, j-k_2) \cdot W(k_1, k_2)\text{; } \quad n_1 \geq k_1 \text{ and } n_2 \geq k_2,
\end{equation}
where Y is the output, commonly known as the feature map. The CNN includes not only convolutional layers, but also sub-sampling layers, to decrease the dimensionality of feature maps, as well as one or more fully connected layers. These fully connected layers play a key role in making the ultimate classification based on the output of the convolutional and sub-sampling layers.

\section{ConvNext Model Architecture}
\label{appendix:convnext architecture}
 \convnext \citep{liu2022convnet} architecture is based on the modified \texttt{ResNet} \citep{he2016deep} architecture, with the stem cell replaced by a ``patchify" layer similar to vision transformers \citep{dosovitskiy2021an}. While the standard \texttt{ResNet} stem cell uses a kernel of 7 x 7  with a stride of 2, followed by a max pool to downsample the image by a factor of 4, the ``patchify" layer contains a 4 x 4 non-overlapping convolution, which is less aggressive than the vision transformers' (14 x 14) or (16 x 16) convolutions. The stride here refers to the number of pixels that the kernel $W$ moves across the input image in each step, and max pooling is a downsampling technique where only a maximum value is selected from a window of feature map, discarding all other values. 
 \begin{figure}[ht!]
  \centering 
  \includegraphics[height=7cm, width=1\textwidth]{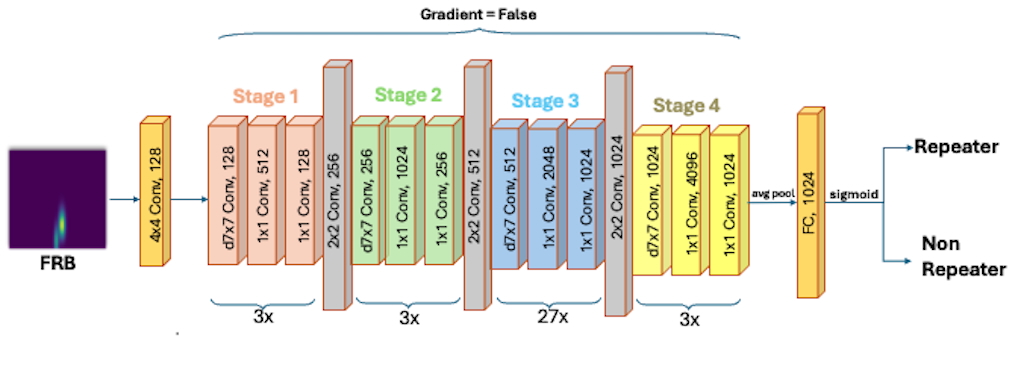} 
  \caption{ConvNext architecture with weights in all stages frozen except ``patchify" layer (the first layer in orange)  and the fully connected layer (the last layer in orange) during fine tuning. The weights are not updated during the fine-tuning for the layers labeled as gradient false. Layers with labels starting with ``d" denote depthwise separable convolution layers followed by the kernel size. ``FC" is the fully connected layer obtained by average pooling (i.e. taking the average values from each feature map window) of the final CNN layer. The output is taken through the sigmoid function \citep{dubey2022activationfunctionsdeeplearning}, which gives a heuristic probability value of an FRB to be in one of the sub-classes.}
  \label{fig:pretrained_convnext}
\end{figure}

\convnext also employs depth-wise convolution \citep{chollet2017xceptiondeeplearningdepthwise} instead of standard convolution, to reduce computational cost. Despite incorporating the ``patchify" layer from vision transformers, \convnext remains a purely convolutional neural network without any transformer or attention mechanism, yet it outperforms some of the vision transformers \citep{liu2022convnet}. The schematic diagram of the \convnext model architecture used for this study is shown in Figure \ref{fig:pretrained_convnext}.

\section{Transfer Learning}
\label{appendix:transfer learning}
Transfer learning is a method for estimating model weights obtained from training on a different data set and implementing them on a related, specific task or data set. This method can maximize efficiency and accuracy by leveraging the knowledge acquired from the pre-trained model \citep{zhuang2020comprehensivesurveytransferlearning}. The \convnext model trained on an original data set is referred to as the pre-trained model, and adjusting the model weights on our own data set is referred to as fine tuning. We used the model pre-trained on the ImageNet-1K \citep{deng2009imagenet} data set from PyTorch's \citep{paszke2019pytorch} torchvision\footnote{\url{https://pytorch.org/vision/stable/models.html}} library \citep{torchvision2016}, with pretraining top 1\% accuracy of $\sim84\%$, $\sim8.8M$ parameters and labeled as \texttt{ConvNext\_Base}. 

\section{Integrated Gradient}
\label{appendix:integrated gradient}
 Integrated gradient (IG) is a  technique that can be used for assessing the relevance of features in the model's decision-making proces \citep{sundararajan2017axiomatic}. Mathematically, an IG is defined as:
\begin{equation}
\label{eq: ig}
    IG_i(x) = (x_i - x_i') \times \int_{\alpha=0}^{1} \frac{\partial F(x' + \alpha \times (x - x'))}{\partial x_i} \, d\alpha,
\end{equation}
where:
\begin{itemize}
    \item \( IG_i(x) \) represents the integrated gradient for the \(i\)-th feature,
    \item \( x \) denotes the actual input, and \( x' \) is a baseline input,
    \item \( F(x) \) is the model’s output function,
    \item \( \alpha \) is a scaling factor that interpolates between the baseline input \( x' \) and the actual input \( x \) and
    \item \( \frac{\partial F(x)}{\partial x_i} \) is the gradient of the model’s output with respect to the \(i\)-th input feature.
\end{itemize}

Equation (\ref{eq: ig}) computes the contribution of each feature (pixel) by integrating the gradients of the model output function along a path from the baseline (black image or 0 pixel value) to the input. These contributions are then visualized as a heatmap and superimposed onto the original input image known as an overlaid integrated gradient, which provides an interpretable explanation of the model's behavior.
}

\input{main.bbl}


\end{document}

%% file: preamble.tex
\PassOptionsToPackage{nosplitbox, valign}{adjustbox}
\usepackage[utf8]{inputenc}

\usepackage{amsmath,amssymb}
\usepackage{array}

\usepackage{multirow}
\usepackage{makecell}
\usepackage{xspace}

\usepackage{xcolor}

\usepackage{booktabs} 
\usepackage{xurl}

\received{2025 September 5}
\revised{2025 December 18}
\accepted{2025 December 29}
\published{2026 February 6}

\newcommand{\nfrbtot}{4545\xspace}
\newcommand{\nfrbrep}{981\xspace}
\newcommand{\nfrbnorep}{3564\xspace}
\newcommand{\nrepeater}{83\xspace}

\newcommand{\cfrb}{CHIME/FRB\xspace}

\newcommand{\catone}{Catalog~1\xspace}
\newcommand{\cattwo}{Catalog~2\xspace}

\newcommand{\fitburst}{\texttt{fitburst}\xspace}
\newcommand{\convnext}{\texttt{ConvNext}\xspace}

\newcommand{\descbox}[1]{\parbox[t]{0.4\linewidth}{\raggedright #1}}

\def\lapp{\ifmmode\stackrel{<}{_{\sim}}\else$\stackrel{<}{_{\sim}}$\fi}
\def\gapp{\ifmmode\stackrel{>}{_{\sim}}\else$\stackrel{>}{_{\sim}}$\fi}

\newcommand{\revieweredit}[1]{{\color{black}#1}}
\newcommand{\referee}[1]{{\color{black}#1}}

%% file: authors.tex
\author[0009-0008-6166-1095]{Bikash Kharel}
    \thanks{The \texttt{morphofrb} framework and source code are available at \url{https://github.com/kharelb/morphofrb}}
    \affiliation{Department of Physics and Astronomy, West Virginia University, PO Box 6315, Morgantown, WV 26506, USA }
    \affiliation{Center for Gravitational Waves and Cosmology, West Virginia University, Chestnut Ridge Research Building, Morgantown, WV 26505, USA}
    
\author[0000-0001-8384-5049]{Emmanuel Fonseca}
    \affiliation{Department of Physics and Astronomy, West Virginia University, PO Box 6315, Morgantown, WV 26506, USA }
    \affiliation{Center for Gravitational Waves and Cosmology, West Virginia University, Chestnut Ridge Research Building, Morgantown, WV 26505, USA}

\author[0000-0002-1800-8233]{Charanjot Brar}
\affiliation{NRC Herzberg Astronomy and Astrophysics, 5071 West Saanich Road, Victoria, BC V9E2E7, Canada}

\author[0009-0004-4176-0062]{Afrokk Khan} 
\affiliation{Department of Physics, McGill University, 3600 rue University, Montr\'eal, QC H3A 2T8, Canada}
\affiliation{Trottier Space Institute, McGill University, 3550 rue University, Montr\'eal, QC H3A 2A7, Canada}

\author[0000-0003-4584-8841]{Lluis Mas-Ribas} 
\affiliation{Department of Astronomy and Astrophysics, University of California, Santa Cruz, 1156 High Street, Santa Cruz, CA 95060, USA}

\author[0009-0008-7264-1778]{Swarali Shivraj Patil} 
\affiliation{Department of Physics and Astronomy, West Virginia University, PO Box 6315, Morgantown, WV 26506, USA }
\affiliation{Center for Gravitational Waves and Cosmology, West Virginia University, Chestnut Ridge Research Building, Morgantown, WV 26505, USA}

\author[0000-0002-7374-7119]{Paul Scholz} 
\affiliation{Department of Physics and Astronomy, York University, 4700 Keele Street, Toronto, ON MJ3 1P3, Canada}
\affiliation{Dunlap Institute for Astronomy and Astrophysics, 50 St. George Street, University of Toronto, ON M5S 3H4, Canada}

\author[0000-0003-2631-6217]{Seth Robert Siegel} 
\affiliation{Perimeter Institute for Theoretical Physics, 31 Caroline Street N, Waterloo, ON N25 2YL, Canada}
\affiliation{Department of Physics, McGill University, 3600 rue University, Montr\'eal, QC H3A 2T8, Canada}
\affiliation{Trottier Space Institute, McGill University, 3550 rue University, Montr\'eal, QC H3A 2A7, Canada}

\author[0000-0002-9761-4353]{David C. Stenning} 
\affiliation{Department of Statistics and Actuarial Science, 8888 University Dr W, Burnaby, BC V5A 1S6, Canada}